\documentclass[aps,pre,reprint,groupedaddress]{revtex4-1}
\usepackage[dvipdfmx]{graphicx}
\usepackage{color}

\begin{document}


\title{Controlled motion of Janus particles \\
in periodically 
phase-separating binary fluids}


\author{Takeaki Araki and Shintaro Fukai}
\affiliation{Department of Physics, Kyoto University, Kyoto 606-8502, Japan}


\def\be{\begin{equation}}
\def\en{\end{equation}}
\def\bea{\begin{eqnarray}}
\def\ena{\end{eqnarray}}

\def\p{\partial}
\def\ep{\epsilon}
\def\gs{\gtrsim}
\def\ls{\lesssim} 
\def\ve{\varepsilon}
\def\n{\nabla}
\def\d{\delta}

\newcommand{\av}[1]{\langle{#1}\rangle}
\newcommand{\AV}[1]{\bigg \langle{#1}\bigg \rangle}
\newcommand{\bi}[1]{\mbox{\boldmath$#1$}}
\newcommand{\pp}[2]{\frac{\partial {#1}}{\partial {#2}}}
\newcommand{\ppp}[3]{{\bigg(}\frac{\partial {#1}}{\partial {#2}}{\bigg )}_{#3}}
\newcommand{\pppd}[3]{{\bigg(}\frac{d {#1}}{d {#2}}{\bigg )}_{#3}}
\newcommand{\pppm}[3]{{\bigg(}\frac{\delta{#1}}{\delta{#2}}{\bigg )}_{#3}}
\newcommand{\ten}[1]{\stackrel{\leftrightarrow}{\bi{#1}}}
\newcommand{\dis}[1]{[{#1}]_-^+}
\newcommand{\di}[1]{\nabla\cdot{{#1}}}
\newcommand{\digra}[2]{\nabla\cdot{#1}\nabla{#2}}

\date{\today}

\begin{abstract}
We numerically investigate
the propelled motions of a Janus particle in a 
periodically phase-separating binary fluid mixture. 
In this study, the surface of the particle tail prefers
one of the binary fluid components and the particle head is neutral 
in the wettability. 
During the demixing period, the more wettable phase is selectively 
adsorbed to the particle tail. 
Growths of the adsorbed domains induce the hydrodynamic flow 
in the vicinity of the particle tail, and 
this asymmetric pumping flow drives the particle toward the particle 
head. 
During the mixing period, 
the particle motion almost ceases because 
the mixing primarily occurs via diffusion and 
the resulting hydrodynamic flow is negligibly small. 
Repeating this cycle unboundedly moves the Janus particle toward the head. 
The dependencies of the composition and the repeat frequency 
on the particle motion are discussed. 
\end{abstract}

\maketitle

\section{Introduction}

Self-propelled motions of micro- and nano-particles 
have attracted much 
interest from a wide range of viewpoints. 
They will provide us with important applications, such as 
nanomachines and drug delivery
\cite{Paxton_JACS_2004,Paxton_ACIE_2006,
Dreyfus_Nature_2005,
Howse_PRL_2007,
Laocharoensuk_AN_2008,
Sundararajan_NL_2008,
Jiang_PRL_2010,Kapral_JCP_2013}. 
Recent, focus has been on their collective dynamics 
because they 
are very fascinating in a growing field of non-equilibrium physics, 
{\it i.e.}, active matter 
\cite{Viscek_PRL_1995,Helding_RPM_2001,Marchetti_RPM_2013,Ramaswamy_ARCMP_2010,Cates_RPP_2012}. 
Self-propelled particles use some energy or nutrients
to generate the self-propulsion force. 
For example, biological molecules, such as ATPase and myosin, 
convert chemical energy to mechanical motion through chemomechanical coupling 
\cite{Schliwa_Nature_2003}. 
In non-biological systems, Marangoni effect 
can induce spontaneous 
motions of liquid droplets \cite{Nagai_PRE_2005,Yabunaka_JCP_2012}.

Janus particles, which have heterogeneous surface properties, 
are often employed as artificial self-propelled systems 
\cite{Golestanian_NJP_2007,Ebbens_SM_2010}. 
For example, the self-propelled motions 
are modelled by asymmetric nanoparticles partially coated with platinum. 
The catalytic decomposition of hydrogen peroxide, 
which occurs selectively on the Pt-surface, drives the nanoparticles 
\cite{Paxton_JACS_2004,Paxton_ACIE_2006,
Laocharoensuk_AN_2008,Golestanian_PRL_2005,Reddy_KARJ_2014}. 
Interfacial phoretic effects 
\cite{Anderson_ARFM_1989,Gangwal_PRL_2008} 
are another possible mechanism of 
micro-swimmers \cite{Golestanian_NJP_2007}. 
Jiang {\it et al.} demonstrated that 
a Janus particle can create 
an asymmetric temperature gradient around it in a defocused laser beam. 
The induced gradients lead to spontaneous 
drift motions of the Janus particles \cite{Jiang_PRL_2010}. 
Also, local heating by illumination light induces active motions 
of Janus particles in a binary mixture of lower critical solution 
temperature \cite{Volpe_SM_2011,Buttinoni_JPCM_2012}. 

Besides the thermo- and diffusio-phoresis 
motions \cite{Jiang_PRL_2010,Volpe_SM_2011,Buttinoni_JPCM_2012}, 
Janus particles can move in phase-separating binary mixtures 
because of the coupling between the wetting and phase separation 
\cite{Cahn_JCP_1977,deGennes_RPM_1985}. 
The phase separation of binary fluid mixtures has been well studied
\cite{Binder_PRL_1974,Siggia_PRA_1979,Onuki_book_2002}. 
During the later stage of the phase separation, 
the domain patterns grow with time. 
The particles in the phase-separating mixtures 
are trapped in one of the phases or at the interfaces. 
Even when Brownian motions and external forces 
are absent, 
the particles move with 
the resulting coarsening of the domain patterns 
\cite{Tanaka_PRL_1984,Ginzburg_PRE_1999,
Araki_PRE_2006,Cates_SM_2008}. 
Because the Janus particles have asymmetric wettability, 
we expect that their motions also become asymmetric. 
The direction of the asymmetric motion will more or less depend 
on the particle direction. 
Janus particles with 
two distinct wettabilities 
are occasionally used as surfactants 
to stabilize the phase-separated domains 
\cite{Binks_Lang_2001,Glaser_Lang_2006,Huang_SM_2012}. 
In the final stage of the phase separation, 
the particle motions will be frozen.

In this article, we demonstrate a possible mechanism 
of spontaneous motions of a Janus particle 
in periodically phase-separating mixtures. 
By continually varying the temperature or pressure 
slightly above and below the transition point, 
one can cause periodic processes of phase separation and mixing
\cite{Onuki_PTP_1982,Onuki_PRL_1982,Joshua_PRL_1985,Tanaka_PRL_1995}. 
By resetting the binary mixtures to the one-phase state, 
we expected that we could continuously propel the particle. 
Here, we examine this expectation by means of numerical simulations. 
The dependencies of 
the particle motion on 
the average composition and the duration of 
the cycle are discussed. 

In Sec.~\ref{sec:model}, we explain our numerical model, which 
is based on fluid particle dynamics 
method \cite{Tanaka_PRL_2000,Tanaka_CES_2006}.
Numerical results are shown and are discussed in Sec.~\ref{sec:results}. 
We summarize our study and discuss some remarks in Sec.~\ref{sec:summary}.

\section{Model and simulation}
\label{sec:model}

\subsection{Free energy functional}

We consider the case, in which a spherical Janus particle 
is suspended in a binary mixture. 
The surface of the particle is heterogeneous in wettability. 
For numerical simulations, dealing with the particle 
in a continuous manner is convenient. 
We express it with a smooth shape 
function as \cite{Tanaka_PRL_2000,Tanaka_CES_2006}, 
\bea
\psi_{\bi{R}}(\bi{r})=\frac{1}{2}
\left\{\tanh\left(\frac{a-|\bi{r}-\bi{R}|}{d}\right)+1\right\}.
\ena
Here, $\bi{r}$ is the coordinate in a lattice space and $\bi{R}$ 
is the position of the particle in an off-lattice space. 
$a$ is the radius of the particle and $d$ represents the 
width of the smooth interface. 
In the limit of $d\rightarrow 0$, 
$\psi$ is unity and zero in the interior and exterior of the particle, respectively. 
We also define the surface distribution as $\psi_{\rm s}=|\n \psi_{\bi{R}}|$. 

The free energy function comprises 
two parts as \cite{Araki_PRE_2006,Araki_JPCM_2008,Furukawa_PRL_2013},
\bea
\mathcal{F}=\mathcal{F}_{\rm mix}+\mathcal{F}_{\rm sur}. 
\label{eq:F}
\ena
The first part $\mathcal{F}_{\rm mix}$ is the mixing free energy given by 
\bea
&&\mathcal{F}_{\rm mix}\{\phi,\bi{R}\}\nonumber\\
&&=
\frac{T}{v_0}
\int d\bi{r}\left[f_{\rm BW}(\phi)+\frac{C}{2}|\n\phi|^2
+\frac{\chi_{\rm p}}{2}\psi_{\bi{R}} (\phi-\phi_0)^2\right],
\label{eq:F_mix} 
\ena
where $\phi(\bi{r})$ is the local concentration 
of the first component of the mixture. 
$T$ is the temperature and we set the Boltzmann constant to unity. 
$v_0$ is the molecular volume. 
The coefficient of the gradient term 
$C$ is related to the interface tension 
and is of the order of $v_0^{2/3}$ \cite{Onuki_book_2002}. 
$f_{\rm BW}(\phi)$ is the Bragg--Williams type of the mixing free energy as 
\bea
f_{\rm BW}(\phi)
=\phi\ln \phi+(1-\phi)\ln(1-\phi)+\chi\phi(1-\phi), 
\label{eq:f_BW}
\ena
where $\chi$ is the interaction parameter between 
two fluid components. 
Under the mean field approximation, 
$\chi=2$ and $\phi=0.5$ give the critical point. 
The third term in the integrand of 
Eq.~(\ref{eq:F_mix}) is introduced to prevent the solvent from penetrating 
into the particle \cite{Araki_PRE_2006}. 
$\chi_{\rm p}$ and $\phi_0$ are the control parameters. 

\begin{figure}[h]
\centering
\includegraphics[width=0.45\textwidth,bb=0 0 645 294]{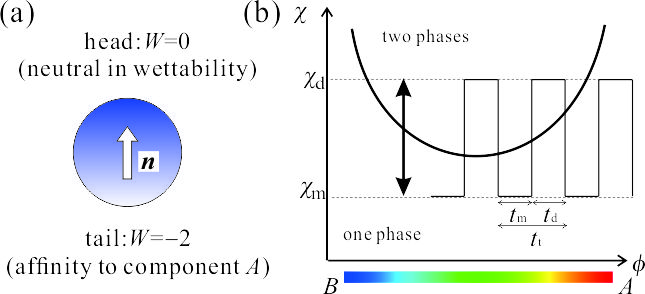}
\caption{
(a) 
A sketch of a spherical particle with a heterogeneous 
surface. 
The orientation is described by a unit vector $\bi{n}$. 
We set $W_0=-1$ and $W_1=1$ in Eq.~(\ref{eq:wettability}), 
so that we have $W=0$ and $W=-2$ at the two poles. 
We refer to the pole with $W=0$ as ``head" and the other as 
``tail". 
The particle head is neutral in wettability and the tail prefers 
component $A$. 
(b) 
A typical phase diagram of a binary fluid mixture. 
We change the $\chi$ parameter by using a square wave function of $t$. 
Phase separation occurs during the period with $\chi=\chi_{\rm d}$. 
However, when $\chi=\chi_{\rm m}$, the phase-separated 
domains are mixed. 
}
\label{fig1}
\end{figure}

$\mathcal{F}_{\rm sur}$ is the surface free energy, which is given by
\bea
&&\mathcal{F}_{\rm sur}\{\phi,\bi{R},\bi{n}\}\nonumber\\
&&=\frac{Td}{v_0}\int d\bi{r}\psi_{\rm s}
(\bi{r})\{\phi(\bi{r})-\phi_{\rm s}\}W(\bi{r}-\bi{R},\bi{n}), 
\ena
where $\bi{n}$ is the unit vector toward the orientation of the 
particle. 
$W$ represents the heterogeneity of the particle wettability. 
As illustrated in Fig.~\ref{fig1}(a), we use 
\bea 
W(\bi{r}-\bi{R},\bi{n})=W_0+W_1 \bi{n}
\cdot \frac{\bi{r}-\bi{R}}{|\bi{r}-\bi{R}|}, 
\label{eq:wettability}
\ena
where $W_0$ and $W_1$ are the material constants 
for the wetting. 
If $W<0$, the component of larger $\phi$ tends to wet the 
surface \cite{Cahn_JCP_1977,deGennes_RPM_1985}. 
In the early stages of phase separation,  
the surface of $W\ne 0$ largely influences 
the pattern formation (see below).

\subsection{Time development equations}

The hydrodynamic equation for the flow field $\bi{v}$ is given by 
\bea
&&\rho\left(\pp{}{t}+\bi{v}\cdot\n\right)\bi{v}\nonumber\\
&&=-\phi\n\frac{\delta \mathcal{F}}{\delta \phi}+\bi{f}+\n \times \bi{g} 
-\n p +\n \bi{\Sigma}. 
\label{eq:NS}
\ena
Here $\rho$ is the material density. 
In this work, we assume that all the materials have the same density. 
$\bi{\Sigma}$ is the viscous stress tensor, which is given by 
\bea
\bi{\Sigma}=\eta \{\n\bi{v}+(\n\bi{v})^T\}. 
\label{eq:viscosity} 
\ena
In the spirit of fluid particle dynamics (FPD), we assume 
that the viscosity $\eta$ depends on the particle distribution as
\cite{Tanaka_PRL_2000,Tanaka_CES_2006}, 
\bea 
\eta(\bi{r})=\eta_0+\Delta \eta \psi_{\bi{R}}(\bi{r}), 
\ena
where $\eta_0$ is the viscosity of the binary fluid and $\Delta \eta$ 
is the viscosity difference between the solvent and the particles. 
In the limit of $\Delta \eta\rightarrow \infty$, the particles 
will behave as solid particles \cite{Tanaka_CES_2006}. 
$p$ is a part of pressure, which 
imposes the incompressibility condition: $\n\cdot\bi{v}=0$. 

$\bi{f}$ is the force field stemming from the particle interactions and 
is given by, 
\bea
\bi{f}=-\frac{\psi_{\bi{R}}
(\bi{r})}{\Omega}\frac{\partial \mathcal{F}}{\partial \bi{R}}, 
\ena
where $\Omega$ is the effective volume defined 
as $\Omega=\int d\bi{r}\psi_{\bi{R}}(\bi{r})$. 
It is approximated as 
$\Omega\approx 4\pi a^3/3$ in three-dimensional systems. 
$\bi{g}$ originates from the torque acting on the particle, 
and is given by 
\bea
\bi{g}=
-\frac{\psi_{\bi{R}}(\bi{r})}{\Omega}
\bi{n}\times \frac{\partial \mathcal{F}}{\partial \bi{n}}.
\ena

The particle motions are caused by the hydrodynamic flow 
and their kinetics are described as 
\bea 
\frac{d}{dt}\bi{R}&=&\bi{V},\\
\bi{V}&=&\frac{1}{\Omega}\int d\bi{r}\psi_{\bi{R}}(\bi{r})\bi{v}, \\
\frac{d}{dt}\bi{n}&=&\frac{1}{\Omega}
\left( \int d\bi{r}\psi_{\bi{R}}(\bi{r})\n\times \bi{v}\right)\times \bi{n}. 
\ena
Furthermore, the time development equation of the concentration field is 
\bea
\pp{}{t}\phi=-\n \cdot (\phi\bi{v})
+\n\cdot L(\psi)\n\frac{\delta \mathcal{F}}{\delta \phi}+
\n\cdot\bi{\zeta}, 
\ena
where $L(\psi)$ is the kinetic coefficient, and in which we set $L(\psi)=L_0(1-\psi)$ 
to eliminate the flux inside the particles. 
$L_0$ is the kinetic coefficient of the bulk mixture. 
$\bi{\zeta}$ represents the thermal fluctuation satisfying the 
fluctuation--dissipation relation. 
In this model, the diffusion flux does not contribute to the particle motion 
directly. 
In each state, the total free energy, including the kinetic energy, 
should decrease with time. 
Its temporal change is described in Appendix~\ref{sec:app1}. 
Our model can be applied to many-particle systems, 
where Janus particles behave as surfactants 
\cite{Binks_Lang_2001,Glaser_Lang_2006,Huang_SM_2012}. 
We hope that we will report our studies on them elsewhere 
in the near future.

\subsection{Numerical simulations}

We numerically solve the above equations 
using the Maker and Cell method with staggered grid \cite{Harlow_PF_1965}. 
We discretize the space by $d$ 
and set $v_0=d^3$ and $C=d^2$. 
Also, we set the particle radius to $a=6d$. 
The simulation box is 
a three-dimensional system ($64^3$) 
with periodic boundary 
conditions. 
The time increment is $0.005\,t_0$, 
where $t_0$ is a typical diffusion time defined by $t_0=d^2T/L_0$. 
For the wettability, we set $W_0=-1$ and $W_1=1$ in this study. 
Hence, the head of the particle is neutral in the wettability 
and 
the tail prefers the component of large $\phi$ 
(see Fig.~\ref{fig1}(a)). 
Hereafter, we describe the more and less wettable components 
as $A$ and $B$, respectively. 
The viscosity parameters in Eq.~(\ref{eq:viscosity}) 
are $\eta_0=\rho L_0/T$ 
and $\Delta \eta=49\eta_0$. 
In Eq.~(\ref{eq:F_mix}), we set $\phi_0$ to be equal to the average 
composition, $\av{\phi}$, and $\chi_{\rm p}=20$. 
Because Reynold numbers in colloidal systems are very small, 
we iterate to integrate Eq.~(\ref{eq:NS}) 
without updating $\bi{R}$, $\bi{n}$, and 
$\phi$ until $|\rho(\partial/\partial t+\bi{v}\cdot\n)\bi{v}|$ becomes 
less than $10^{-3}\eta_0 L_0/(T_0d^3)$.
The intensity of the thermal fluctuation 
is given by 
$\av{\zeta_i(\bi{r},t)\zeta_j(\bi{r}',t')}
=0.05TL(\psi)
\delta(\bi{r}-\bi{r}')\delta(t-t')\delta_{ij}$, 
where $i$ and $j$ stand for $x$, $y$ and $z$.

\subsection{Periodic phase separation}

To induce periodic phase separation, 
we change the $\chi$ parameter uniformly in space 
using square wave functions of time 
(see Fig.~\ref{fig1}(b)). 
$t_{\rm m}$ and $t_{\rm d}$ denote the 
durations for mixing and demixing, respectively. 
We also define $t_{\rm t}=t_{\rm m}+t_{\rm d}$. 
For simplicity, because we consider 
mixtures near the phase-separation point, 
we assume that the other physical parameters 
are constant, independent of $\chi$. 
In durations of $mt_{\rm t}\le t 
< mt_{\rm t}+t_{\rm m}$, 
$\chi$ is fixed at $\chi_{\rm m}$ 
below the critical point. 
Here, $m$ is an integer. 
The phase-separated domains will be mixed during 
these times. 
In $mt_{\rm t}+t_{\rm m}\le t< (m+1)t_{\rm t}$, 
we retain $\chi$ at $\chi=\chi_{\rm d}$, 
above the critical point, 
so that phase separation proceeds in the bulk. 
Throughout the simulations presented in this article,
we fix the $\chi$ parameters to $\chi_{\rm d}=2.7$ and 
$\chi_{\rm m}=1.3$. 
In the demixing periods with $\chi_{\rm d}$, 
the coexistence concentrations of $A$- and $B$-rich phases 
are $\phi\cong 0.893$ and $0.107$, respectively. 
The correlation length, also referred to as the interface thickness, is
$\xi_{\rm d}=\sqrt{C/(\chi-\chi_{\rm c})}\cong 1.20d$. 
However, 
during the mixing periods with $\chi_{\rm m}$, 
the correlation length is given by 
$\xi_{\rm m}=\sqrt{C/\{2(\chi_{\rm c}-\chi)\}}\cong 0.845d$.

\section{Results and discussions}
\label{sec:results}

\subsection{Domain patterns}

\begin{figure}[h]
\centering
\includegraphics[width=0.4\textwidth,bb= 0 0 1311 1984]{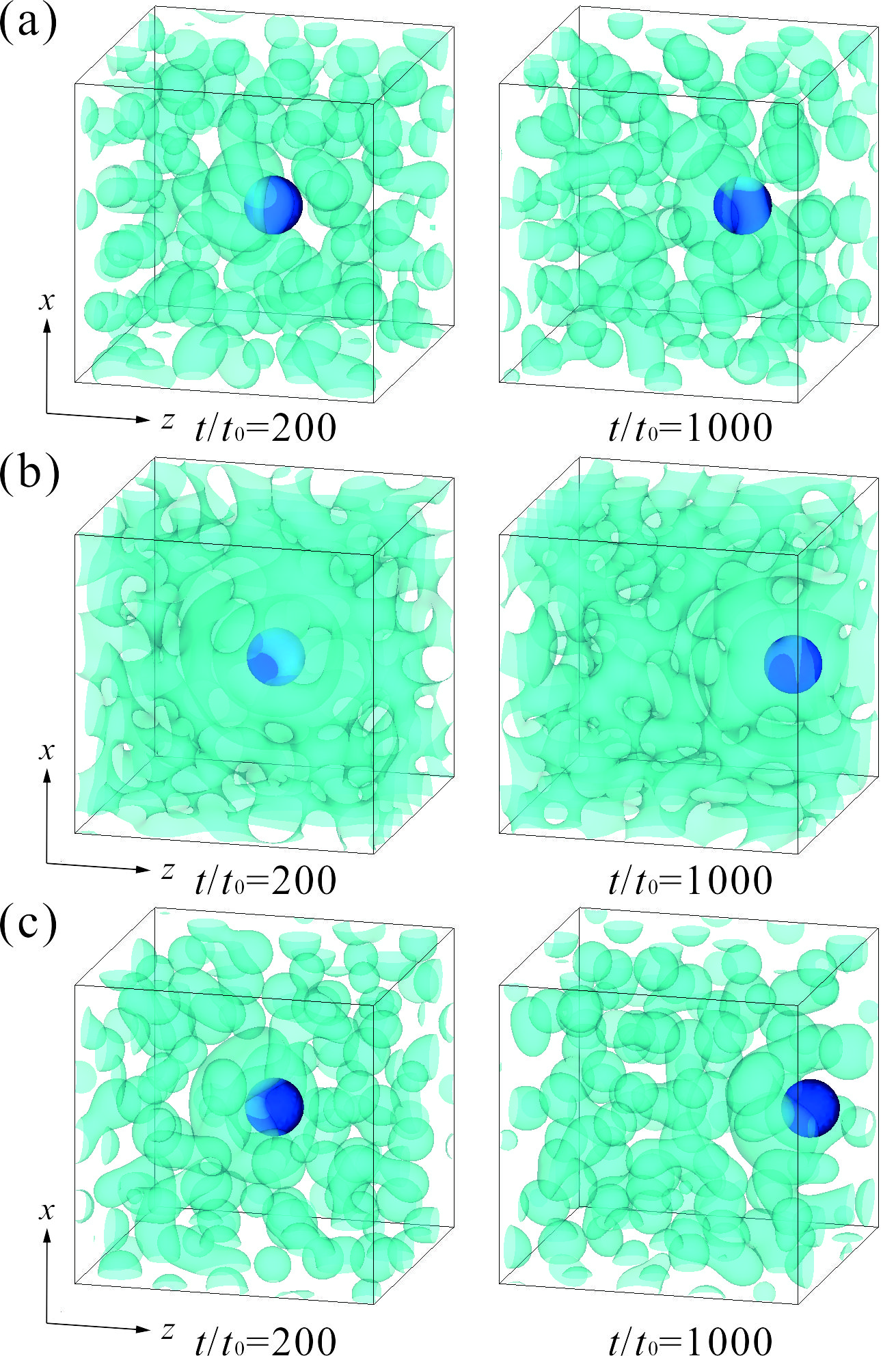}
\caption{
Snapshots of a Janus particle in periodically 
phase-separating binary fluids. 
The average concentrations of the more wettable phase 
are 
(a) $\av{\phi}=0.3$, 
(b) $\av{\phi}=0.5$, and 
(c) $\av{\phi}=0.7$.}
\label{fig2}
\end{figure}

Figure \ref{fig2} shows 
snapshots of the domain patterns 
in the case of $t_{\rm m}=t_{\rm d}=100\,t_0$. 
The snapshots were obtained at $t=200\,t_0$ and 
$t=1000\,t_0$, which correspond to the times 
when the first and fifth demixing periods have 
just finished, respectively. 
The average concentrations, $\av{\phi}$, of component $A$ are 
(a) 0.3, (b) 0.5, and (c) 0.7. 
In the demixing periods with $\chi_{\rm d}$,  
the volume fractions of the $A$-rich phase 
are (a) $24.6\%$, (b) $50\%$ and (c) $75.4\%$, respectively. 
The light blue surfaces represent 
the isosurfaces of $\phi=0.5$ and 
the dark blue sphere represents the Janus particle. 
At $t=0$, the particle is directed toward the $z$-axis, 
{\it i.e.}, $\bi{n}(t=0)=(0,0,1)$. 
In each case, 
we observed that the particle follows the background flows, which are caused by the interface 
tensions during the phase separations. 
The mechanisms of the particle motions are discussed below. 

\subsection{Particle trajectories}

\begin{figure}[h]
\centering
\includegraphics[width=0.45\textwidth, bb= 0 0 1310 536]{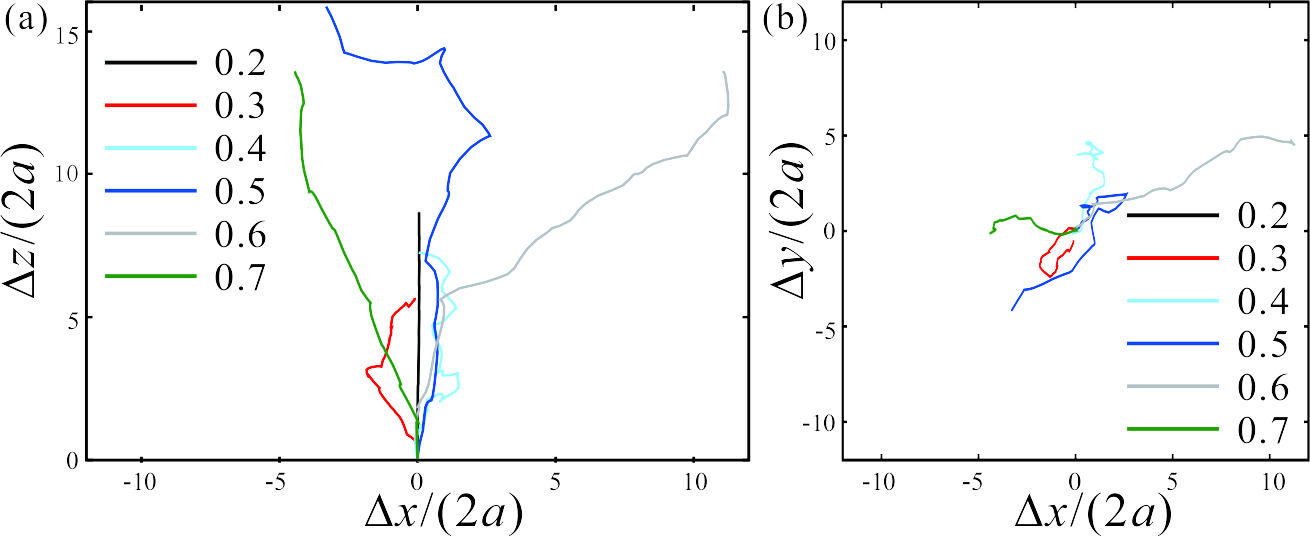}
\caption{
The trajectories of the Janus particle in the periodically 
phase separating mixtures from $t=0$ to $t=10^4\,t_0$. 
The trajectories are projected on the (a) $x$--$z$ and 
(b) $x$--$y$ planes. 
At $t=0$, the particle is directed toward the $z$-axis.
The time intervals are $t_{\rm d}=t_{\rm m}=500\,t_0$. 
}
\label{fig3}
\end{figure}

Figure \ref{fig3} shows typical trajectories of the 
Janus particle from $t=0$ to $t=10^4\,t_0$ 
in the binary mixtures of different concentrations. 
The time intervals are $t_{\rm m}=t_{\rm d}=500\,t_0$. 
The trajectories show that the particle tends to move directionally 
toward the particle head. 
Because the particle surface has a heterogeneous affinity to the 
components, 
the phase separation proceeds asymmetrically around the particle. 
This asymmetry of the phase separation dynamics may cause the 
directional motion of the Janus particle. 
Interestingly, the particle moves toward its head 
in all the mixtures. 
Figure~\ref{fig3} also 
indicates that the trajectories are not completely straight. 
The degree of the directionality and the particle speed 
depend on parameters such as 
the average concentrations and the time intervals. 

\begin{figure}[h]
\centering
\includegraphics[width=0.3\textwidth, bb= 0 0 590 849]{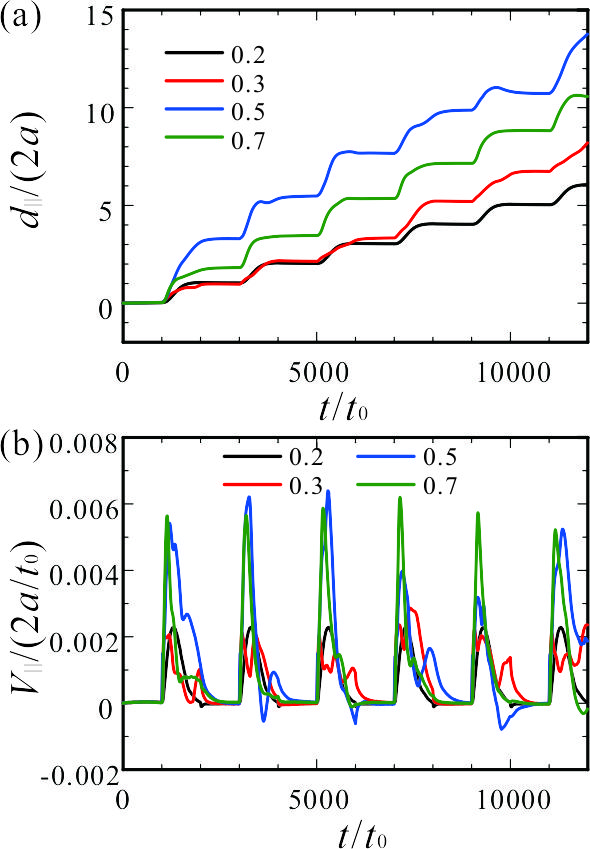}
\caption{
The time developments of (a) the particle displacements 
$d_\parallel$ 
and the particle velocity $V_\parallel$ 
toward the particle head. 
The time intervals are $t_{\rm d}=t_{\rm m}=10^3\,t_0$. 
The average concentrations are 
$\av{\phi}=0.2,\, 0.3,\,0.5$ and $0.7$. 
The hatched regions indicate the mixing periods. 
}
\label{fig4}
\end{figure}

Figures~\ref{fig4}(a) and (b) show 
the temporal changes in the 
trajectory length and velocity toward the head. 
Since the particle orientation, $\bi{n}$, changes with time, 
the trajectory length $d_{\parallel}$ 
and the velocity $V_{\parallel}$ toward the particle orientation 
are 
calculated as 
\bea
d_{\parallel}(t)&=&\int_0^t V_\parallel(t')dt',\\
V_{\parallel}(t)&=&\bi{V}(t)\cdot \bi{n}(t). 
\ena
Positive and negative values of $V_{\parallel}$ represnt the 
forward and backward motions of the particle position, respectively. 
In Fig.~\ref{fig4}, the time intervals are set to 
$t_{\rm m}=t_{\rm d}=10^3\,t_0$. 
Here, each curve was obtained from one 
simulation run. 
We simulated nine average concentrations 
from $\av{\phi}=0.1$ to $0.9$. 
In the mixtures of $\av{\phi}\le 0.1$ and $\av{\phi}\ge 0.8$, 
we did not observe any drastic motion of the particle; 
hence, their curves have not been included in Fig.~\ref{fig4}. 
Also, we did not plot the curves for the mixtures of $\av{\phi}=0.4$ and $\av{\phi}=0.6$ because they essentially demonstrate behaviors 
similar to those with the symmetric mixture $\av{\phi}=0.5$. 

In Fig.~\ref{fig4}(a), the trajectory lengths indicate stepwise motions. 
The particle is almost fixed in the mixing periods. 
However, during the demixing periods,  
the particle shows forward displacements. 
By repeating these cyclic motions, 
the particle continuously propels in a periodically phase-separating 
binary mixture. 
The displacement in each cycle is of the order of the particle 
diameter. 
Although the onsets of the motion in the demixing periods 
are not clearly seen in Fig.~\ref{fig4}, 
the detailed analyses indicate 
that the particle does not start moving simultaneously 
with the quenching into the demixing states. 
It moves most largely after 
a certain incubation time $t_{\rm i}$, 
which is discussed later.  
 
The cyclic 
behaviors are also clearly displayed in the particle velocity. 
After the initial incubation time 
in each demixing period, 
the particle velocity shows large positive values. 
After this transient deterministic motion, 
the velocity decreases gradually with 
some fluctuations.  
In particular, it can have negative values in more symmetric mixtures, with
$\av{\phi}=0.5$. 
A similar stepwise motion is observed in 
a system where a Janus particle with metallic surfaces 
is trapped 
at a liquid--air interface \cite{Reddy_KARJ_2014}. 
The particle's stepwise motion is due to spontaneous cyclic bursts of 
bubbles. 
In our system, the stepwise motion is due to the controlled 
changes of the interaction parameter.

\begin{figure}[h]
\centering
\includegraphics[width=0.3\textwidth, bb= 0 0 580 823]{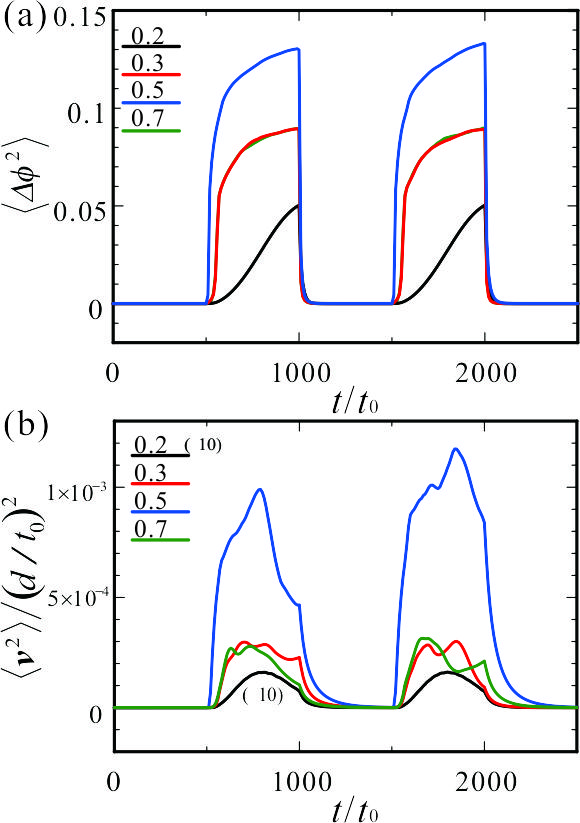}
\caption{
The time developments of 
(a) the 
average-concentration differences $\av{\Delta \phi^2}$ and 
the velocity intensity $\av{\bi{v}^2}$. 
The time intervals are $t_{\rm d}=t_{\rm m}=500\,t_0$. 
In (b), 
the curve for $\av{\phi}=0.2$ is magnified tenfold. 
The average concentrations are 
$\av{\phi}=0.2,\, 0.3,\,0.5$ and $0.7$. 
The hatched regions indicate the mixing periods. 
}
\label{fig5}
\end{figure}

\begin{figure}[h]
\centering
\includegraphics[width=0.3\textwidth, bb= 0 0 609 1261]{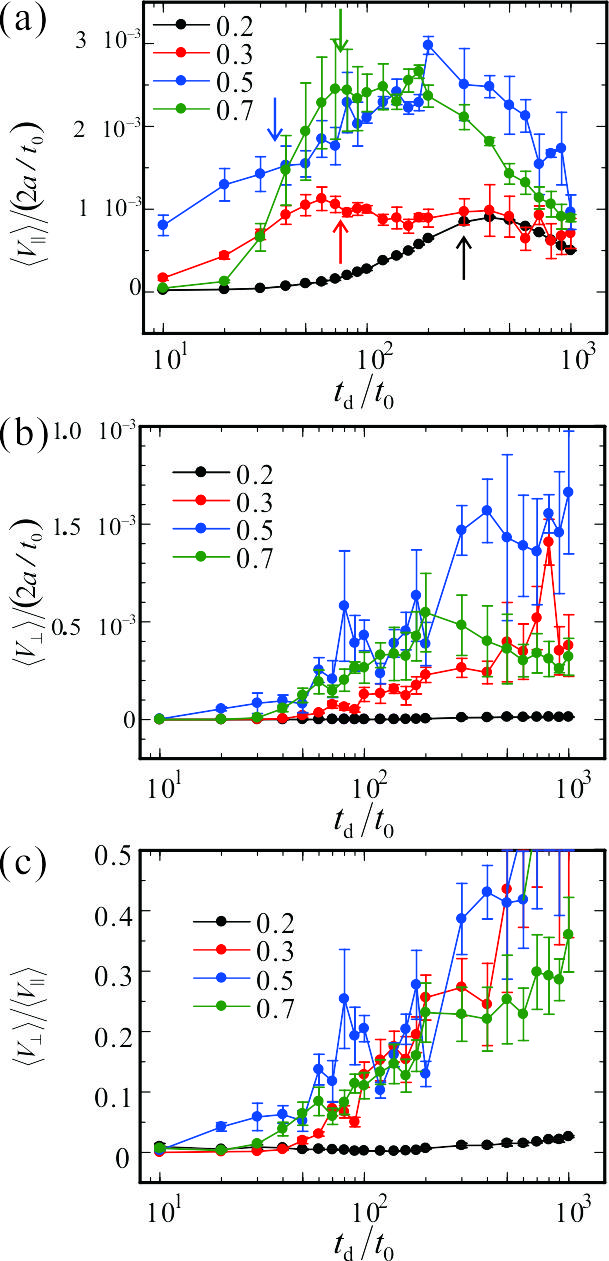}
\caption{
Plots of the 
averaged velocity (a) toward the particle head $\av{V_\parallel}$ 
and (b) perpendicular to the orientation $\av{V_\perp}$, 
with respect to the time intervals. 
The arrows in (a) indicate the durations of the 
early stage $t_{\rm e}$. 
The average concentrations are 
$\av{\phi}=0.2,\, 0.3,\,0.5$ and $0.7$. 
The error bars represent the standard deviation of the 
particle motions. 
(c) 
The ratio of the perpendicular speed $\av{V_\perp}$ to 
the parallel speed $\av{V_\parallel}$. 
}
\label{fig6}
\end{figure}

Figures~\ref{fig5}(a) and (b) show the 
temporal changes of the average-concentration differences 
and the velocity intensities. 
They are calculated as 
\bea
\av{\Delta \phi^2}&=&\Omega_{\rm t}^{-1}\int d\bi{r}(\phi-\av{\phi})^2,\\
\av{\bi{v}^2}&=&\Omega_{\rm t}^{-1}\int d\bi{r}\bi{v}^2, 
\ena
where $\Omega_{\rm t}$ is the system volume. 
In the early stages of the demixing periods, 
the phase separation starts via diffusion of the components. 
As shown in Fig.~\ref{fig5}, 
the hydrodynamic flow develops 
simultaneously with the phase separation, 
and thus, it is small during the early stages 
\cite{Tanaka_PRL_1998,Tanaka_EPL_2000}. 
Because the particle is transported by the background flow, 
the incubation time of the particle motion $t_{\rm i}$ in 
Fig.~\ref{fig4} corresponds to the 
duration of the early stage of the phase separation, $t_{\rm e}$. 
Here, $t_{\rm e}$ depends 
on the average concentration. 
This is because the growth rate of the concentration field 
depends on the average concentration. 
Inside the spinodal regime, the phase separation 
proceeds via spinodal decomposition. 
As the average concentration approaches the spinodal 
points ($\av{\phi}=0.5\pm 0.255$ for $\chi_{\rm d}=2.7$), 
the growth rate is decreased to zero. 
However, in the binodal regime, 
the phase separation occurs via the nucleation of the 
droplets of the minority phase. 
The nucleation rate is also decreased to zero as $\av{\phi}$ 
approaches the equilibrium concentration. 
In both the processes, 
more symmetric mixtures are more unstable and 
the durations of the early stages are shortened. 
This may suggest that the particle moves faster in the symmetric 
mixtures because the incubation time, $t_{\rm i}$, 
during which the particle is at rest, is reduced. 
We define the duration of the early stage $t_{\rm e}$ 
as $\av{\Delta \phi^2(mt_{\rm t}+t_{\rm m}+t_{\rm e})}
=\av{\Delta \phi_{\rm eq}^2}/2$. 
Here, $\Delta\phi_{\rm eq}$ is the concentration difference 
in the equilibrium demixing state. 
From the simulation results, 
we obtain $t_{\rm e}=34.6\,t_0$ for $\av{\phi}=0.5$, 
$t_{\rm e}\approx 74.0\,t_0$ for $\av{\phi}=0.3$ and $0.7$, and 
$t_{\rm e}\approx  292\,t_0$ for $\av{\phi}=0.2$. 
These are indicated by the arrows in Fig.~\ref{fig6}(a).

Figs.~\ref{fig6}(a) and (b) show plots of the averaged 
speeds toward the particle head, $\av{V_\parallel}$, 
and perpendicular to it, $\av{V_\perp}$, 
as a function of the 
time interval. 
Here, we set $t_{\rm m}=t_{\rm d}$ for simplicity. 
The parallel and perpendicular velocity in the $m$-th cycle 
are defined as 
\bea
V_{\parallel m }
&=&\frac{1}{t_{\rm t}}
\int_{mt_{\rm t}}^{(m+1)t_{\rm t}}
V_\parallel(t)dt,\\
V_{\perp m}&=&
\frac{1}{t_{\rm t}}
\int_{m t_{\rm t}}^{(m+1)t_{\rm t}}
|\bi{V}(t)\times \bi{n}(t)|dt.
\ena
From these, 
the averaged velocities $\av{V_\parallel}$ and $\av{V_\perp}$ 
are obtained with averaging $M=10$ cycles as
\bea
\av{V_X}=\frac{1}{M}\sum_{m=1}^MV_{Xm}, 
\ena
where $X$ refers to $X=\parallel$ and $\perp$. 
The error bars in Figs.~\ref{fig6}(a) and (b) represent the standard 
deviations of $V_{\parallel m}$ and $V_{\perp m}$. 
The ratio of $\av{V_\perp}$ to $\av{V_\parallel}$ is plotted 
in Fig.~\ref{fig6}(c). 

In Fig.~\ref{fig6}(a), 
each curve of $\av{V_\parallel}$ is non-monotonic with maxima. 
The maxima peaks suggest that we can choose efficient time intervals for 
propelling the particle. 
As discussed above, 
the hydrodynamic flow has not developed yet in the early stage 
of the demixing periods. 
In the cases of small time intervals, 
the phase-separating times are too short for the hydrodynamic flows to develop sufficiently; 
hence, the particle is not dragged, 
implying that we have to maintain the system in the phase-separated state 
till the conclusion of the early stage $t_{\rm e}$. 
However, for large time intervals, 
the average speed along the orientation becomes low. 
In the mixtures of $\av{\phi}=0.2,\,0.3$, and $0.7$, 
the parallel speed becomes largest 
at approximately $t_{\rm d}\cong t_{\rm e}$. 
Thus, larger time intervals are not needed for 
propelling the particle with high speeds. 
However, in the symmetric mixture $(\av{\phi}=0.5)$, 
the parallel speed becomes largest around 
$t_{\rm d}\approx 200\,t_0$, which is 
larger than the duration of an early stage $t_{\rm e}$. 
This is attributed to large background flows emerging from 
other domains near the particle, as discussed below. 

\begin{figure}[h]
\centering
\includegraphics[width=0.3\textwidth, bb= 0 0 604 846]{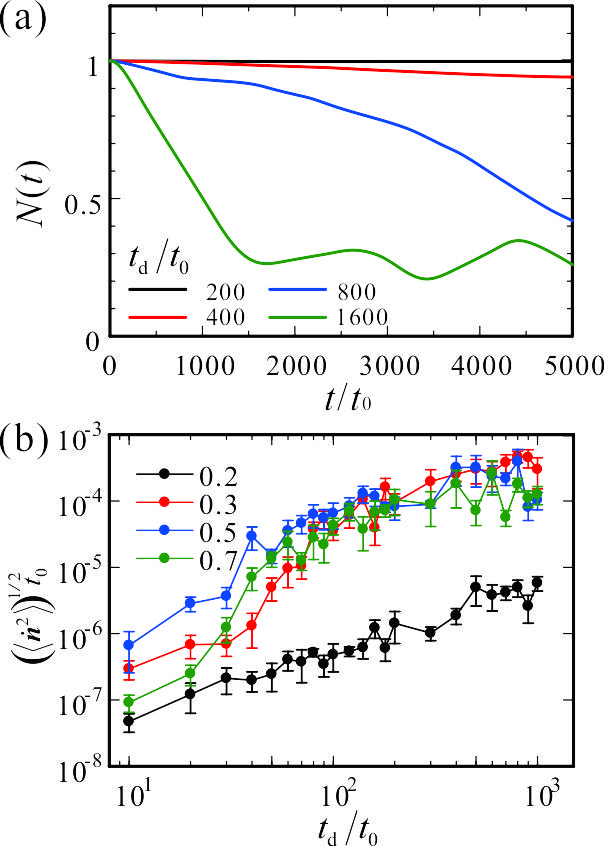}
\caption{
(a) The autocorrelation function of the orientation $N(t)$. 
The average concentration was constant at $\av{\phi}=0.3$, whereas 
the time intervals changed. 
(b) 
The dependence of the change rate of the particle orientation 
$(\av{\dot{\bi{n}^2}})^{1/2}$ on the time interval. 
}
\label{fig7}
\end{figure}

As has been shown in Fig.~\ref{fig3}, 
we observe the particle fluctuations, which are 
characterized by the perpendicular motion $\av{V_\perp}$ in 
Fig.~\ref{fig6}(b). 
The particle changes its orientation and the resulting 
direction of the particle motion.  
Thus, similar to the $\av{V_\perp}$ changes, the temporal changes of the orientation are also 
considered as a measure 
of the particle fluctuation. 
Fig.~\ref{fig7}(a) demonstrates the 
autocorrelation of the orientation vector. 
It is calculated as 
\bea
N(t)=\frac{1}{t_{\rm max}}\int_0^{t_{\rm max}} 
dt' \bi{n}(t')\cdot\bi{n}(t+t'), 
\ena
where we set $t_{\rm max}=5\times 10^3t_0$. 
In Fig.~\ref{fig7}(a), we plot the autocorrelation for $\av{\phi}=0.3$ 
as a typical example. 
They decrease with time 
indicating that the memory of the orientation is gradually 
lost. 
Figure~\ref{fig7}(a) shows that the decay rate is increased with 
increasing $t_{\rm d}$. 
The autocorrelations for different $\av{\phi}$ behave in the same way. 
In Fig.~\ref{fig7}(b), we plot the
change rate of the orientation $(\av{\dot{\bi{n}}^2})^{1/2}$, 
which is calculated for $M=10$ as 
\bea
\av{ \dot{\bi{n}}^2}&=&\frac{1}{M}\sum_{m=0}^{M-1}\dot{\bi{n}}_m^2.\\
\dot{ \bi{n}}_m
&=&\frac{1}{t_{\rm t}}\{\bi{n}((m+1)t_{\rm t})-\bi{n}(mt_{\rm t})\}.
\ena

As shown in Figs.~\ref{fig6}(b) and \ref{fig7}(b), 
both $\av{V_\perp}$ and $(\av{\dot{\bi{n}}^2})^{1/2}$ 
are large for large values of $t_{\rm d}$, in contrast 
to those for small $t_{\rm d}$. 
In the demixing periods, 
the phase separation proceeds throughout the bulk. 
The fluctuations of the particle motions and orientation stem from 
the hydrodynamic flows 
accompanied by the spontaneous growth of the domains surrounding 
the particle. 
These background flows are independent of the particle; 
thus, they disturb the particle motion and change its orientation. 
Therefore, if we use large time intervals, 
the particle motion is likely to deviate from the straight line 
along the initial orientation $\bi{n}(t=0)$.
As suggested in Figs.~\ref{fig6} and \ref{fig7}, 
the ratios $\av{V_\perp}/\av{V_\parallel}$ and 
$(\av{\dot{\bi{n}}}^2)^{1/2}$ increased slightly with $t_{\rm d}$. 
Also, in this sense, the large time interval is not preferred for the 
controlled propulsions.

Figure~\ref{fig6}(a) shows that the propelling speed is higher 
in the symmetric mixtures ($\av{\phi}=0.5$) 
than in the asymmetric mixtures. 
However, the fluctuations of the particle motion 
are also large as shown 
in Figs.~\ref{fig6}(b) and \ref{fig7}(b). 
Therefore, the symmetric mixtures are not suitable 
for keeping the straight line motions. 
The particle moves largely; however, its motion easily loses 
the directionality with time. 
Furthermore, in the asymmetric mixtures, 
the fluctuations of the particle motion and 
the orientation are relatively 
small for the preferred $t_{\rm d}$; thus, 
the asymmetric mixtures are more suitable to control the particle 
motions. 
We consider the mechanisms of the propelled motions in 
the two types of asymmetric mixtures separately.

\subsection{The wettable component-rich mixture}

\begin{figure}[h]
\centering
\includegraphics[width=0.4\textwidth,bb= 0 0 1300 1402]{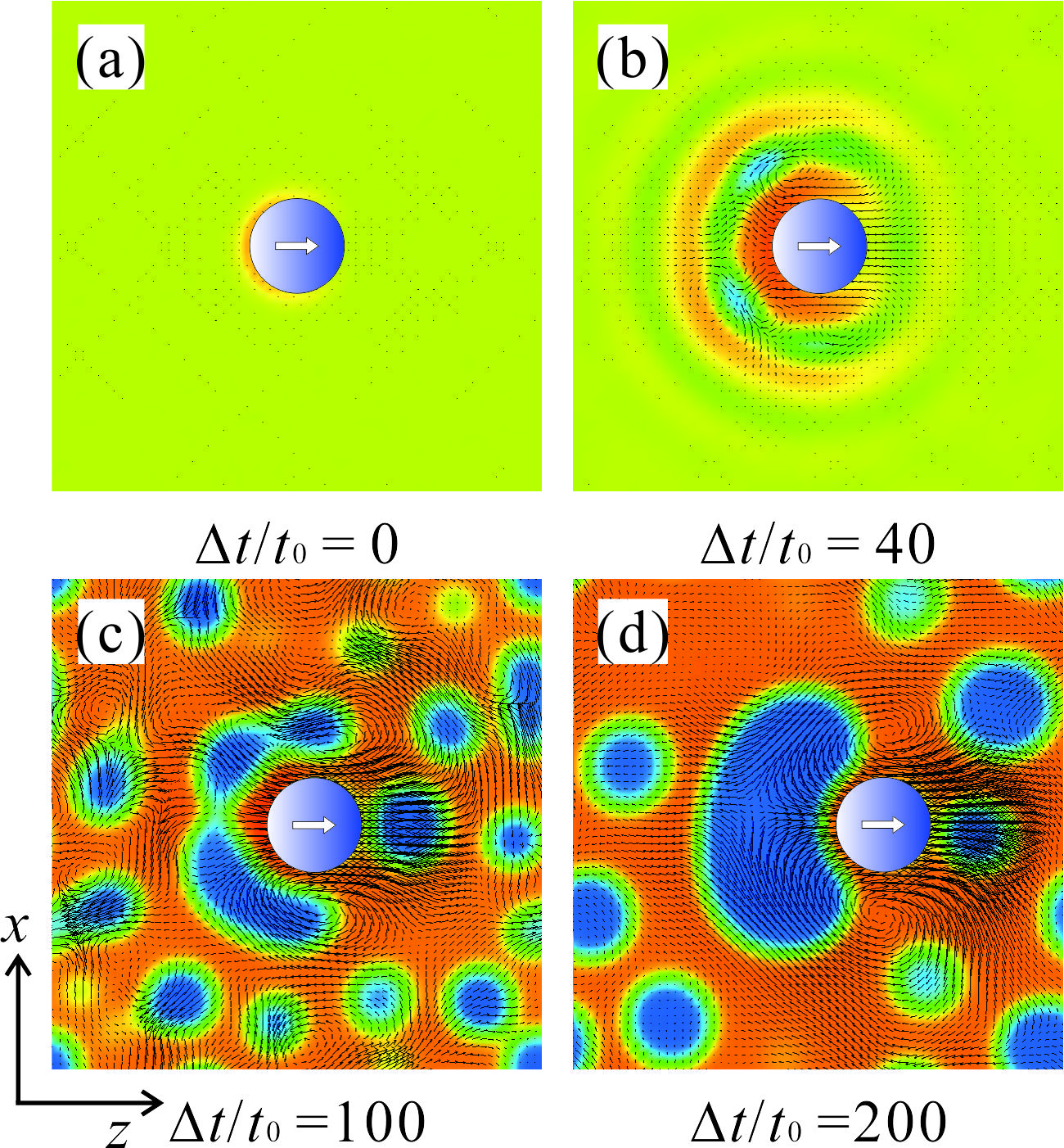}
\caption{
Snapshots of typical pattern evolution around the particle 
in the $A$-rich mixtures ($\av{\phi}=0.7$). 
The arrows show the flow field. 
The red and blue domains are the $A$- and $B$-rich phases, 
respectively. 
The time intervals are $t_{\rm d}=t_{\rm m}=300\,t_0$. 
}
\label{fig8}
\end{figure}

Figure~\ref{fig8} shows the patterns of the evolutions of the 
concentration and flow field 
in the mixture of $\av{\phi}=0.7$. 
As shown in Fig.~\ref{fig3}, 
the particle moves rather straightforward. 
In the early stage of the phase separation during the 
demixing periods, 
the $A$-rich phase wets the half-portion 
of the particle and a wetting layer is formed on it. 
Because of the directional fluxes of the $A$-component toward 
the surface, the $A$-component is depleted near the outside of 
this first wetting layer. 
This accumulated layer structure is 
similar to 
the oscillating profiles of the concentration field 
near a flat wall or a particle with a homogeneous surface
\cite{Karim_Macro_1999,Tanaka_EPL_2000}. 
To compensate for the depleted region, 
non-spherical droplets of the $B$-rich phase are formed as shown 
in Fig.~\ref{fig8}(b).
The droplets grow near the surface via coagulation and 
coalescence. 
Because the processes of the coagulation and coalescence 
occur asymmetrically near the particle tail, 
the associated hydrodynamic flow likely 
pushes the particle toward the particle head. 
This process is shown in Figs.~\ref{fig8}(c) and (d). 
During the phase separation, many small droplets of the 
$B$-rich phase are also formed in the bulk. 
They grow to their typical sizes with time via 
coalescence and coagulation, or evaporation and condensation processes 
\cite{Binder_PRL_1974,Onuki_book_2002}. 
Around coalescing droplets, other hydrodynamic flows are 
induced, which disturb the particle motion. 
The disturbing flows are relatively weaker; 
thus, the particle motion remains straightforward, 
in contrast to that in the symmetric mixtures. 

The mixture of $\av{\phi}=0.8$ should be phase-separated 
at equilibrium, when $\chi=\chi_{\rm d}(=2.7)$. 
However, because the nucleation rate is very small, the time intervals we employed ($t_{\rm d}\le 10^3\,t_0)$ were 
not enough to induce the phase separation. 
Thus, we did not observe any motion of the Janus particle 
in the mixture of $\av{\phi}\ge 0.8$.

\subsection{The wettable component-deficient mixture}

\begin{figure}[h]
\centering
\includegraphics[width=0.4\textwidth, bb= 0 0 1311 1380]{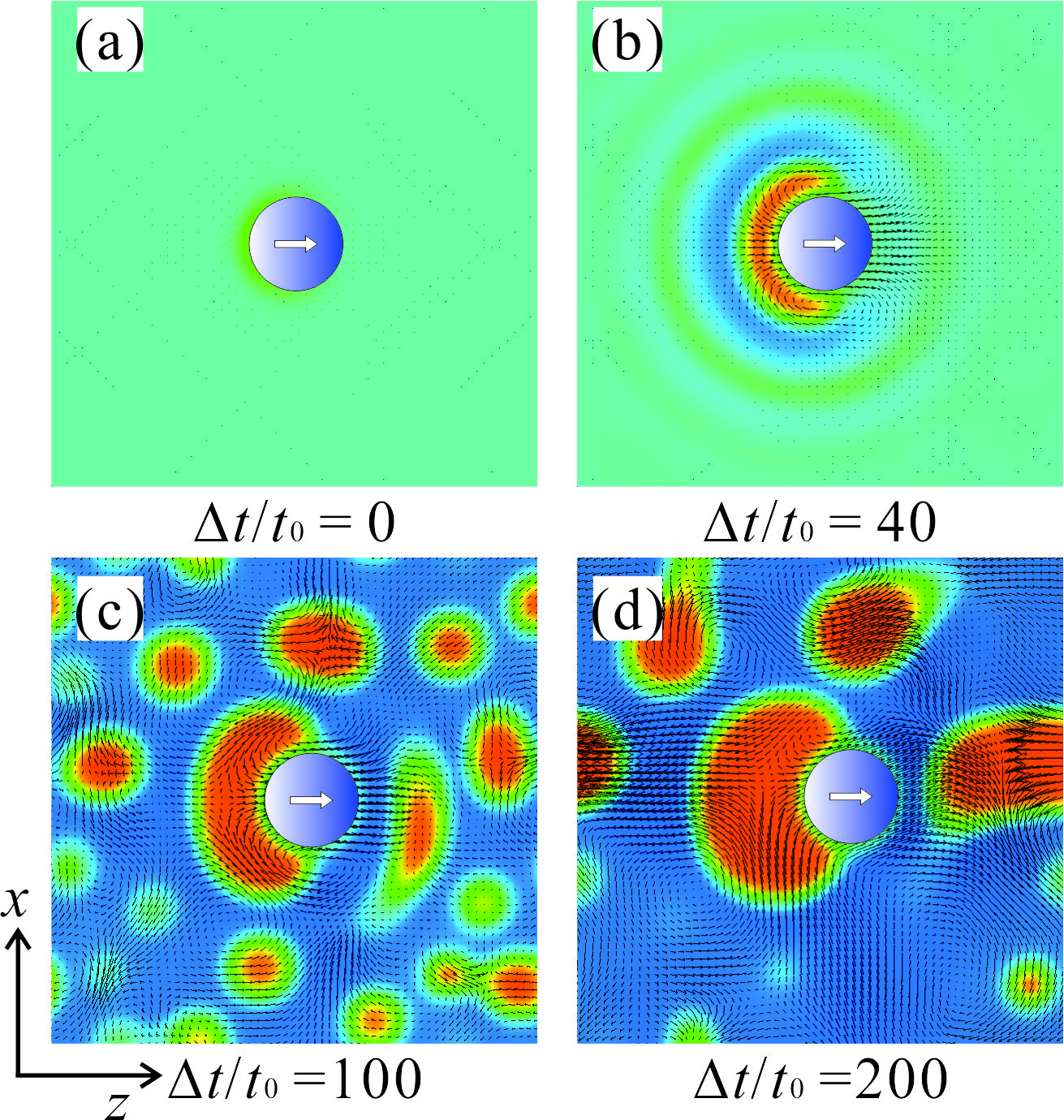}
\caption{
Snapshots of typical pattern evolutions around the particle 
in the $B$-rich mixtures ($\av{\phi}=0.3$). 
The arrows show the flow field. 
The red and blue domains are the $A$- and $B$-rich phases, 
respectively. 
The time intervals are $t_{\rm d}=t_{\rm m}=300\,t_0$. 
}
\label{fig9}
\end{figure}

The motions in 
the $B$-rich mixtures 
are highly directed toward the particle orientation as shown 
in Figs.~\ref{fig3} and \ref{fig6}. 
Figure~\ref{fig9} shows 
the pattern evolutions of the 
concentration and flow field 
in the mixture of $\av{\phi}=0.3$. 
In the early stages of the phase separation, 
a cap-shaped domain of the $A$-rich phase emerges from the 
wettable portion of the particle surface. 
It covers a large amount of the interfacial area, 
although its volume is not so large. 
Then, it tends to change its shape to a sphere 
for reducing the interfacial energy 
after the early stage of the phase separation. 
This process induces a pumping hydrodynamic flow
around the tail of the Janus particle 
and the resulting pumping flow 
pushes the particle toward the head as shown in 
Figs.~\ref{fig9}(c) and (d). 
The decay time of the localized pumping flow, $t_{\rm h}$, 
is estimated as $t_{\rm h}\approx \eta a/\sigma$, 
where $\sigma$ is the interface tension.

The mixtures of $\av{\phi}=0.3$ and $0.7$ have 
the same stability for the phase separation in the bulk. 
Because the tail of the particle prefers component $A$, 
this asymmetry leads to the difference in the particle motions 
between the $\av{\phi}=0.3$ and the $\av{\phi}=0.7$ cases. 
Figure~\ref{fig6}(a) shows that the highest parallel speed in 
in the $\av{\phi}=0.7$ mixture is approximately twice of 
that in the $\av{\phi}=0.3$ case. 
However, 
Figure~\ref{fig6}(c) indicates that 
the ratio $\av{V_\perp}/\av{V_\parallel}$ 
in the $\av{\phi}=0.7$ case is also approximately twice of 
that in the $\av{\phi}=0.3$ case 
at the maximum parallel speed. 
Also, Figure~\ref{fig6}(a) shows that 
the standard deviations of $V_{\rm \perp m}$ 
for the $\av{\phi}=0.3$ mixture 
are smaller than those for the $\av{\phi}=0.7$ case. 
The particle is suggested to move more steadily and smoothly 
in the $\av{\phi}=0.3$ mixture. 
Thus, we conclude that the $B$-rich mixtures are more preferred to induce 
more straight motions. 

The stability for the bulk phase-separation in the mixture 
of $\av{\phi}=0.2$ is the same as that in the $\av{\phi}=0.8$ case. 
However, we have not observed any motions in the $\av{\phi}=0.8$ mixture, 
although the particle moves straightforward in the $\av{\phi}=0.2$ case. 
This difference suggests that the domain formation in the mixture of 
$\av{\phi}=0.2$ is attributed to the heterogeneous nucleation 
at the particle surface \cite{Winter_PRL_2009}.  
As the average concentration approaches the binodal line, 
the thermal nucleation rate in the bulk is strongly 
decreased \cite{Onuki_book_2002}. 
However, the rate of heterogeneous nucleation 
on the wetting surface is large enough to induce it during 
our demixing periods. 
Because the number of the surrounding droplets is decreased, 
particle motion becomes more straightforward 
in more asymmetrically $B$-rich mixtures 
(see the case of $\av{\phi}=0.2$ in Fig.~\ref{fig6}). 

In both cases, 
the hydrodynamic flow around the particle tail pushes the 
particle toward the head. 
In this sense, our particle motion may be categorized 
to a pusher in the active matter field \cite{Marchetti_RPM_2013}. 
However, more detailed analyses on the flow pattern are 
required before this conclusion can be derived.

\subsection{Roles of mixing periods}

Figures \ref{fig4}(a) and (b) indicate 
that the particle does not show any large changes of the 
position and orientation 
during the mixing periods. 
This is because diffusion dominates the mixing but it does not 
contribute to the particle motion. 
However, here, we should note that 
the mixing periods are very important for resetting 
the binary mixtures. 
As noted above, 
the hydrodynamic flow caused by the interface tension 
is large in the demixing periods, whereas it is small 
in the mixing periods. 
This difference in the hydrodynamic flows present between the mixing and demixing periods induces the continuous 
propelled motion of the Janus particle. 

In the above simulations, we set $t_{\rm m}=t_{\rm d}$ for simplicity. 
However, considering whether this mixing interval 
is sufficient to reset the binary mixtures is important. 
The characteristic length $\ell$ of the 
phase-separated domain increases with time algebraically. 
In the symmetric mixture, a bicontinuous pattern is formed 
and the domain grows obeying \cite{Siggia_PRA_1979}
\bea
\ell(t)\approx c\sigma t/\eta.
\label{eq:ell1}
\ena
However, in the asymmetric mixtures, 
the minority phase forms droplets. 
The droplets grow with time 
via coalescence and coagulation 
as \cite{Binder_PRL_1974,Onuki_book_2002} 
\bea
\ell(t)\approx c'(Tt/\eta)^{1/3}.
\label{eq:ell2}
\ena
Here, $c$ and $c'$ are non-dimensional numbers, 
which depend on the volume fractions. 
Then, the characteristic length at the ends of the demixing periods 
would be given by $\ell(t_{\rm d})$. 
If the volume fraction is quite small, 
the droplets grow via nucleation and growth 
mechanism, and 
the above scaling relation (Eq.~(\ref{eq:ell2}))
is replaced by 
$\ell(t)\propto (L_0\sigma t)^{1/3}$. 
Because its growth exponent is the same as that of Eq.~(\ref{eq:ell2}), 
we consider only cases described by Eq.~(\ref{eq:ell2}) above. 

In the mixing periods, these phase-separated domains 
should be dissolved into 
a homogeneous state via diffusion. 
The diffusion time is estimated as 
$t_{\rm dif}\sim \ell(t_{\rm d})^2/D_{\rm m}$, where 
$D_{\rm m}=2L_0(\chi_{\rm sp}-\chi_{\rm m})/T$ 
is the diffusion constant at $\chi=\chi_{\rm m}$ 
and $\chi_{\rm sp}$ is the interaction parameter at the spinodal point. 
When the time interval of the mixing period is longer than 
the diffusion time, 
the system can be reset for the next demixing period. 
However, if $t_{\rm dif}\ll t_{\rm m}$, 
the mixtures show non-steady states 
\cite{Onuki_PTP_1982,Onuki_PRL_1982,Joshua_PRL_1985,Tanaka_PRL_1995}
and the particle would not move directionally. 
The condition $t_{\rm dif}\ll t_{\rm m}$ is rewritten as 
$t_{\rm m}\gg (T/\eta)^{2/3}t_{\rm d}^{2/3}/D_{\rm m}$ 
for the droplet patterns and 
$t_{\rm m}\gg (\sigma/\eta)^2t_{\rm d}^2/D_{\rm m}$ for 
the bicontinuous patterns. 
If we set $t_{\rm d}\gg (T/\eta)^2D^{-3}_{\rm m}$, 
we have $t_{\rm dif}\ll t_{\rm d}$ in the asymmetric mixtures. 
Therefore, long annealing times for the mixing are not 
required. 
The total time interval $t_{\rm t}=t_{\rm m}+t_{\rm d}$ 
can be reduced to $t_{\rm t}\cong t_{\rm d}$.

\section{Summary and remarks}
\label{sec:summary}

We proposed a possible mechanism of propelled motions of 
Janus particles in periodically phase-separating binary mixtures 
using numerical simulations. 
Because the particle has an asymmetric surface in wettability, 
phase separation proceeds heterogeneously around the particle. 
The resulting asymmetric hydrodynamic flow leads to directional 
motions of the particle. 
After a usual one-step quench, the directional motion slows down and 
will stop eventually.
By changing the interaction parameter, the phase-separated 
system recovers to a one-phase mixing state. 
Therefore, under periodic changes of the interaction parameter, 
we can induce a continuous motion along the particle orientation. 

We found that the particle propels more directionally in asymmetric 
binary mixtures. 
In symmetric mixtures, the hydrodynamic flow emerging from other domains 
surrounding the particle is so large that it disturbs the directional 
motions of the Janus particle. 
We also found that the propelling speed and the directionality 
depend on the frequency 
of the change in the interaction parameter. 
In the cases of short time intervals, 
the hydrodynamic flow does not develop well and 
it cannot drive the particle. 
However, in the cases of long time intervals, 
the hydrodynamic flows from the surrounding domains disturb 
the directional motion. 
We can efficiently move the particle in the intermediate time 
intervals, which are comparable to the characteristic time 
of the spinodal decomposition or nucleations in  
the early stage of the phase separation. 
The pumping hydrodynamic flow that is localized around the particle 
decays with the relaxation time $t_{\rm h}$. 
$t_{\rm h}$ is of the order of $\eta a/\sigma$. 
The interval of the demixing period should be larger than 
$t_{\rm i}+t_{\rm h}$. 
However, the large time interval for the demixing periods 
leads to the loss of the directionality as discussed above. 
Thus, the most efficient time interval would be 
$t_{\rm d}\approx t_{\rm i}+t_{\rm h}$. 
The displacement of the particle in each cycle is of the 
order of the particle diameter. 
Then, the maximum particle speed could be increased to 
$V\sim a/(t_{\rm i}+t_{\rm h})$. 

Unfortunately, 
the simulations performed in this study are limited 
owing to the numerical costs. 
We need to deepen our understanding of the propelled motions of such a Janus particle, and we hope that 
we will present a report on them in the near future. 
We make some critical remarks to improve our study as follows. 

(1) 
In this article, we show only the simulations with 
the particle diameter $2a=12d$. 
Here, $d$ is comparable to the correlation length; 
hence, our particle is rather small. 
In actual phase-separating mixtures, 
such small particles would show drastic 
Brownian motions and the directional motions 
we discovered might be smeared out. 
Our preliminary simulations with larger particles 
indicated that the displacement of the particle in each 
cycle is of the order of the particle size. 
In other words, 
they suggest that the particle speed can be increased 
linearly with its size by employing appropriate temporal 
changes of the interaction parameter.

(2) 
The wettability of our particle changes smoothly 
on the surface (see Eq.~(\ref{eq:wettability})). 
However, an actual Janus particle usually 
has two distinct surfaces 
and the wettability changes abruptly at the equator. 
Instead of Eq.~(\ref{eq:wettability}), 
we performed some simulations with 
an alternate surface function given by 
\bea
W(\bi{r}-\bi{R},\bi{n})
=W_0+W_1\tanh
\left(\frac{1}{d_h}
\bi{n}\cdot\frac{\bi{r}-\bi{R}}{|\bi{r}-\bi{R}|}
\right). 
\label{eq:wettability2}
\ena
Here, 
$d_h$ is introduced to avoid the singularity 
at the equator of the Janus particles. 
A particle described by Eq.~(\ref{eq:wettability2}) with small $d_h$ 
would behave more realistically. 
The preliminary simulations using 
Eq.~(\ref{eq:wettability2}) with $d_h=0.05$ 
demonstrated essentially similar results as 
those using Eq.~(\ref{eq:wettability}). 
This similarity can be derived from the fact that 
our particle motions are induced by the secondary effect 
of the phase separation, {\it i.e.,} the asymmetric 
growth of the hydrodynamic flow around the particle. 
Thus, we consider that our findings are 
robust for the heterogeneous surface pattern. 
To improve the efficiency of the particle motions, 
simulations with a variety of the surface 
structures would be interesting. 

We consider that 
the propelled motions in this article 
are robust also for the details of the free energy 
function. 
We numerically confirmed that the particle moves in the same way 
in binary mixtures described by the Ginzburg--Landau free energy 
instead of Eq.~(\ref{eq:f_BW}).

(3) 
The easiest method to realize our findings in actual systems 
would be a direct observation of them with an optical 
microscope equipped with a temperature 
control hot stage \cite{Tanaka_PRL_1995}. 
However, changing 
the temperature with high frequencies might be experimentally difficult 
because the thermal diffusion constant is finite. 
To induce a high speed propulsion, 
large differences of the temperature 
from the binodal point are preferred 
for both the mixing and demixing periods. 
However, for deep quenches in the demixing periods,  
the incubation times for the phase separation would be 
reduced up to microscopic timescales. 

The pressure control \cite{Shibayama_Macro_2004} 
is considered to be an alternative 
method to induce the continuous propulsions. 
The combination of the temperature control and 
illumination-induced phase separation would also be able 
to induce periodic phase separation with high speeds 
\cite{Volpe_SM_2011,Buttinoni_JPCM_2012}. 
In this study, we employ only the square waves of the 
interaction parameter to induce periodic phase separation. 
We should perform 
more simulations with other types of wave functions 
to find more efficient propulsion schemes. 

(4) 
In our model, particle motion is caused by the hydrodynamic flow. 
However, studying the behaviors of a Janus particle in 
solid mixtures where $\bi{v}=0$ would also be interesting. 
In such mixtures, the particle motion is caused by the thermodynamic 
forces $(\propto \n\delta \mathcal{F}/\delta \phi)$. 
Because the coarsening behavior of the phase separation domain 
pattern depends 
on the fluidity of the mixtures (see Eqs.(\ref{eq:ell1}) and (\ref{eq:ell2})), 
whether our findings are applicable to the solid mixture 
 is not trivial. 

(5) 
Because the intensity of the hydrodynamic flow is proportional to $1/\eta_0$ 
(see Eq.~(\ref{eq:NS})), 
we expected that the particle speed is also proportional to $1/\eta_0$. 
However, our simulations with different solvent viscosities 
indicated 
that $V_\parallel$ is approximately proportional to $1/\eta_0$ in symmetric mixtures, 
whereas it is approximately independent of $\eta_0$ in asymmetric mixtures 
(not shown here). 
This is because the rate-limiting process of the phase separation in asymmetric mixtures is 
the nucleation and growth 
of the minority phase; their rates are dominated 
by the diffusion constant, not the solvent viscosity 
. 
Thus, the results reported in this article are not 
quantitatively universal. 
They will depend on the system parameters such as the solvent viscosity 
and interface tension. 
Although we consider that the propelled motion in periodic 
phase separation can be 
qualitatively realized in actual systems, 
we have to perform more simulations with wide ranges of the 
system parameters. 

For simplicity, 
we also assumed a constant viscosity $\eta_0$ in the surrounding fluids.  
However, the solvent viscosity is generally a function of 
the local composition. 
As noted above, the particle speed depends on the solvent viscosity in 
symmetric mixtures; 
thus, studying the influences of the viscosity 
difference would also be interesting.

\section*{Acknowledgements}

The authors thank H. Tanaka for his helpful comments. 
We started this work following his original ideas on 
this research topic. 
This work was supported by 
KAKENHI (no. 24540433, 23244088 and 25000002) 
and the 
JSPS Core-to-Core Program "International research network 
for non-equilibrium dynamics of soft matter." 
The computational work was performed using the facilities at the 
Supercomputer Center, Institute for Solid State Physics, University of Tokyo. 
This research was also partly supported by CREST, JST.

\bibliography{janus} 

\begin{thebibliography}{50}%
\makeatletter
\providecommand \@ifxundefined [1]{%
 \@ifx{#1\undefined}
}%
\providecommand \@ifnum [1]{%
 \ifnum #1\expandafter \@firstoftwo
 \else \expandafter \@secondoftwo
 \fi
}%
\providecommand \@ifx [1]{%
 \ifx #1\expandafter \@firstoftwo
 \else \expandafter \@secondoftwo
 \fi
}%
\providecommand \natexlab [1]{#1}%
\providecommand \enquote  [1]{``#1''}%
\providecommand \bibnamefont  [1]{#1}%
\providecommand \bibfnamefont [1]{#1}%
\providecommand \citenamefont [1]{#1}%
\providecommand \href@noop [0]{\@secondoftwo}%
\providecommand \href [0]{\begingroup \@sanitize@url \@href}%
\providecommand \@href[1]{\@@startlink{#1}\@@href}%
\providecommand \@@href[1]{\endgroup#1\@@endlink}%
\providecommand \@sanitize@url [0]{\catcode `\\12\catcode `\$12\catcode
  `\&12\catcode `\#12\catcode `\^12\catcode `\_12\catcode `\%12\relax}%
\providecommand \@@startlink[1]{}%
\providecommand \@@endlink[0]{}%
\providecommand \url  [0]{\begingroup\@sanitize@url \@url }%
\providecommand \@url [1]{\endgroup\@href {#1}{\urlprefix }}%
\providecommand \urlprefix  [0]{URL }%
\providecommand \Eprint [0]{\href }%
\providecommand \doibase [0]{http://dx.doi.org/}%
\providecommand \selectlanguage [0]{\@gobble}%
\providecommand \bibinfo  [0]{\@secondoftwo}%
\providecommand \bibfield  [0]{\@secondoftwo}%
\providecommand \translation [1]{[#1]}%
\providecommand \BibitemOpen [0]{}%
\providecommand \bibitemStop [0]{}%
\providecommand \bibitemNoStop [0]{.\EOS\space}%
\providecommand \EOS [0]{\spacefactor3000\relax}%
\providecommand \BibitemShut  [1]{\csname bibitem#1\endcsname}%
\let\auto@bib@innerbib\@empty
\bibitem [{\citenamefont {Paxton}\ \emph {et~al.}(2004)\citenamefont {Paxton},
  \citenamefont {Kistler}, \citenamefont {Olmeda}, \citenamefont {Sen},
  \citenamefont {St.~Angelo}, \citenamefont {Cao}, \citenamefont {Mallouk},
  \citenamefont {Lammert},\ and\ \citenamefont {Crespi}}]{Paxton_JACS_2004}%
  \BibitemOpen
  \bibfield  {author} {\bibinfo {author} {\bibfnamefont {W.~F.}\ \bibnamefont
  {Paxton}}, \bibinfo {author} {\bibfnamefont {K.~C.}\ \bibnamefont {Kistler}},
  \bibinfo {author} {\bibfnamefont {C.~C.}\ \bibnamefont {Olmeda}}, \bibinfo
  {author} {\bibfnamefont {A.}~\bibnamefont {Sen}}, \bibinfo {author}
  {\bibfnamefont {S.~K.}\ \bibnamefont {St.~Angelo}}, \bibinfo {author}
  {\bibfnamefont {Y.}~\bibnamefont {Cao}}, \bibinfo {author} {\bibfnamefont
  {T.~E.}\ \bibnamefont {Mallouk}}, \bibinfo {author} {\bibfnamefont {P.~E.}\
  \bibnamefont {Lammert}}, \ and\ \bibinfo {author} {\bibfnamefont {V.~H.}\
  \bibnamefont {Crespi}},\ }\href@noop {} {\bibfield  {journal} {\bibinfo
  {journal} {J. Am. Chem. Soc.}\ }\textbf {\bibinfo {volume} {126}},\ \bibinfo
  {pages} {13424} (\bibinfo {year} {2004})}\BibitemShut {NoStop}%
\bibitem [{\citenamefont {Paxton}\ \emph {et~al.}(2006)\citenamefont {Paxton},
  \citenamefont {Sundararajan}, \citenamefont {Mallouk},\ and\ \citenamefont
  {Sen}}]{Paxton_ACIE_2006}%
  \BibitemOpen
  \bibfield  {author} {\bibinfo {author} {\bibfnamefont {W.~F.}\ \bibnamefont
  {Paxton}}, \bibinfo {author} {\bibfnamefont {S.}~\bibnamefont
  {Sundararajan}}, \bibinfo {author} {\bibfnamefont {T.~E.}\ \bibnamefont
  {Mallouk}}, \ and\ \bibinfo {author} {\bibfnamefont {A.}~\bibnamefont
  {Sen}},\ }\href@noop {} {\bibfield  {journal} {\bibinfo  {journal} {Angew.
  Chem. Int. Ed.}\ }\textbf {\bibinfo {volume} {45}},\ \bibinfo {pages} {5420}
  (\bibinfo {year} {2006})}\BibitemShut {NoStop}%
\bibitem [{\citenamefont {Dreyfus}\ \emph {et~al.}(2005)\citenamefont
  {Dreyfus}, \citenamefont {Baudry}, \citenamefont {Roper}, \citenamefont
  {Stone},\ and\ \citenamefont {Bibette}}]{Dreyfus_Nature_2005}%
  \BibitemOpen
  \bibfield  {author} {\bibinfo {author} {\bibfnamefont {R.}~\bibnamefont
  {Dreyfus}}, \bibinfo {author} {\bibfnamefont {J.}~\bibnamefont {Baudry}},
  \bibinfo {author} {\bibfnamefont {M.}~\bibnamefont {Roper}, \bibfnamefont
  {M.~L.~Fermigier}}, \bibinfo {author} {\bibfnamefont {H.~A.}\ \bibnamefont
  {Stone}}, \ and\ \bibinfo {author} {\bibfnamefont {J.}~\bibnamefont
  {Bibette}},\ }\href@noop {} {\bibfield  {journal} {\bibinfo  {journal}
  {Nature}\ }\textbf {\bibinfo {volume} {437}},\ \bibinfo {pages} {862}
  (\bibinfo {year} {2005})}\BibitemShut {NoStop}%
\bibitem [{\citenamefont {Howse}\ \emph {et~al.}(2007)\citenamefont {Howse},
  \citenamefont {Jones}, \citenamefont {Ryan}, \citenamefont {Gough},
  \citenamefont {Vafabakhsh},\ and\ \citenamefont
  {Golestanian}}]{Howse_PRL_2007}%
  \BibitemOpen
  \bibfield  {author} {\bibinfo {author} {\bibfnamefont {J.~R.}\ \bibnamefont
  {Howse}}, \bibinfo {author} {\bibfnamefont {R.~A.~L.}\ \bibnamefont {Jones}},
  \bibinfo {author} {\bibfnamefont {A.}~\bibnamefont {Ryan}}, \bibinfo {author}
  {\bibfnamefont {T.}~\bibnamefont {Gough}}, \bibinfo {author} {\bibfnamefont
  {R.}~\bibnamefont {Vafabakhsh}}, \ and\ \bibinfo {author} {\bibfnamefont
  {R.}~\bibnamefont {Golestanian}},\ }\href@noop {} {\bibfield  {journal}
  {\bibinfo  {journal} {Phys. Rev. Lett.}\ }\textbf {\bibinfo {volume} {99}},\
  \bibinfo {pages} {048102} (\bibinfo {year} {2007})}\BibitemShut {NoStop}%
\bibitem [{\citenamefont {Laocharoensuk}\ \emph {et~al.}(2008)\citenamefont
  {Laocharoensuk}, \citenamefont {Burdick},\ and\ \citenamefont
  {Wang}}]{Laocharoensuk_AN_2008}%
  \BibitemOpen
  \bibfield  {author} {\bibinfo {author} {\bibfnamefont {R.}~\bibnamefont
  {Laocharoensuk}}, \bibinfo {author} {\bibfnamefont {J.}~\bibnamefont
  {Burdick}}, \ and\ \bibinfo {author} {\bibfnamefont {J.}~\bibnamefont
  {Wang}},\ }\href@noop {} {\bibfield  {journal} {\bibinfo  {journal} {ASC
  Nano}\ }\textbf {\bibinfo {volume} {5}},\ \bibinfo {pages} {1069} (\bibinfo
  {year} {2008})}\BibitemShut {NoStop}%
\bibitem [{\citenamefont {Sundararajan}\ \emph {et~al.}(2008)\citenamefont
  {Sundararajan}, \citenamefont {Lammert}, \citenamefont {Zudans},
  \citenamefont {Crespi},\ and\ \citenamefont {Sen}}]{Sundararajan_NL_2008}%
  \BibitemOpen
  \bibfield  {author} {\bibinfo {author} {\bibfnamefont {S.}~\bibnamefont
  {Sundararajan}}, \bibinfo {author} {\bibfnamefont {P.~E.}\ \bibnamefont
  {Lammert}}, \bibinfo {author} {\bibfnamefont {A.~W.}\ \bibnamefont {Zudans}},
  \bibinfo {author} {\bibfnamefont {V.~H.}\ \bibnamefont {Crespi}}, \ and\
  \bibinfo {author} {\bibfnamefont {A.}~\bibnamefont {Sen}},\ }\href@noop {}
  {\bibfield  {journal} {\bibinfo  {journal} {NANO Lett.}\ }\textbf {\bibinfo
  {volume} {8}},\ \bibinfo {pages} {1271} (\bibinfo {year} {2008})}\BibitemShut
  {NoStop}%
\bibitem [{\citenamefont {Jiang}\ \emph {et~al.}(2010)\citenamefont {Jiang},
  \citenamefont {Yoshinaga},\ and\ \citenamefont {Sano}}]{Jiang_PRL_2010}%
  \BibitemOpen
  \bibfield  {author} {\bibinfo {author} {\bibfnamefont {H.-R.}\ \bibnamefont
  {Jiang}}, \bibinfo {author} {\bibfnamefont {N.}~\bibnamefont {Yoshinaga}}, \
  and\ \bibinfo {author} {\bibfnamefont {M.}~\bibnamefont {Sano}},\ }\href@noop
  {} {\bibfield  {journal} {\bibinfo  {journal} {Phys. Rev. Lett.}\ }\textbf
  {\bibinfo {volume} {105}},\ \bibinfo {pages} {268302} (\bibinfo {year}
  {2010})}\BibitemShut {NoStop}%
\bibitem [{\citenamefont {Kapral}(2013)}]{Kapral_JCP_2013}%
  \BibitemOpen
  \bibfield  {author} {\bibinfo {author} {\bibfnamefont {R.}~\bibnamefont
  {Kapral}},\ }\href@noop {} {\bibfield  {journal} {\bibinfo  {journal} {J.
  Chem. Phys.}\ }\textbf {\bibinfo {volume} {138}},\ \bibinfo {pages} {020901}
  (\bibinfo {year} {2013})}\BibitemShut {NoStop}%
\bibitem [{\citenamefont {Vicsek}\ \emph {et~al.}(1995)\citenamefont {Vicsek},
  \citenamefont {Czir\'{o}k}, \citenamefont {Ben-Jacob}, \citenamefont
  {Cohen},\ and\ \citenamefont {Shochet}}]{Viscek_PRL_1995}%
  \BibitemOpen
  \bibfield  {author} {\bibinfo {author} {\bibfnamefont {T.}~\bibnamefont
  {Vicsek}}, \bibinfo {author} {\bibfnamefont {A.}~\bibnamefont {Czir\'{o}k}},
  \bibinfo {author} {\bibfnamefont {E.}~\bibnamefont {Ben-Jacob}}, \bibinfo
  {author} {\bibfnamefont {I.}~\bibnamefont {Cohen}}, \ and\ \bibinfo {author}
  {\bibfnamefont {O.}~\bibnamefont {Shochet}},\ }\href@noop {} {\bibfield
  {journal} {\bibinfo  {journal} {Phys Rev. Lett.}\ }\textbf {\bibinfo {volume}
  {75}},\ \bibinfo {pages} {1226} (\bibinfo {year} {1995})}\BibitemShut
  {NoStop}%
\bibitem [{\citenamefont {Helding}(2001)}]{Helding_RPM_2001}%
  \BibitemOpen
  \bibfield  {author} {\bibinfo {author} {\bibfnamefont {D.}~\bibnamefont
  {Helding}},\ }\href@noop {} {\bibfield  {journal} {\bibinfo  {journal} {Rev.
  Mod. Phys.}\ }\textbf {\bibinfo {volume} {73}},\ \bibinfo {pages} {1067}
  (\bibinfo {year} {2001})}\BibitemShut {NoStop}%
\bibitem [{\citenamefont {Marchetti}\ \emph {et~al.}(2013)\citenamefont
  {Marchetti}, \citenamefont {Joanny}, \citenamefont {Ramaswamy}, \citenamefont
  {Liverpool}, \citenamefont {Prost}, \citenamefont {Rao},\ and\ \citenamefont
  {Aditi~Simha}}]{Marchetti_RPM_2013}%
  \BibitemOpen
  \bibfield  {author} {\bibinfo {author} {\bibfnamefont {M.~C.}\ \bibnamefont
  {Marchetti}}, \bibinfo {author} {\bibfnamefont {J.-F.}\ \bibnamefont
  {Joanny}}, \bibinfo {author} {\bibfnamefont {S.}~\bibnamefont {Ramaswamy}},
  \bibinfo {author} {\bibfnamefont {T.~B.}\ \bibnamefont {Liverpool}}, \bibinfo
  {author} {\bibfnamefont {J.}~\bibnamefont {Prost}}, \bibinfo {author}
  {\bibfnamefont {M.}~\bibnamefont {Rao}}, \ and\ \bibinfo {author}
  {\bibfnamefont {R.}~\bibnamefont {Aditi~Simha}},\ }\href@noop {} {\bibfield
  {journal} {\bibinfo  {journal} {Rev. Mod. Phys.}\ }\textbf {\bibinfo {volume}
  {85}},\ \bibinfo {pages} {1143} (\bibinfo {year} {2013})}\BibitemShut
  {NoStop}%
\bibitem [{\citenamefont {Ramaswamy}(2010)}]{Ramaswamy_ARCMP_2010}%
  \BibitemOpen
  \bibfield  {author} {\bibinfo {author} {\bibfnamefont {S.}~\bibnamefont
  {Ramaswamy}},\ }\href@noop {} {\bibfield  {journal} {\bibinfo  {journal}
  {Annu. Rev. Condens. Matter Phys.}\ }\textbf {\bibinfo {volume} {1}},\
  \bibinfo {pages} {323} (\bibinfo {year} {2010})}\BibitemShut {NoStop}%
\bibitem [{\citenamefont {Cates}(2012)}]{Cates_RPP_2012}%
  \BibitemOpen
  \bibfield  {author} {\bibinfo {author} {\bibfnamefont {M.~E.}\ \bibnamefont
  {Cates}},\ }\href@noop {} {\bibfield  {journal} {\bibinfo  {journal} {Rep.
  Prog. Phys.}\ }\textbf {\bibinfo {volume} {75}},\ \bibinfo {pages} {042601}
  (\bibinfo {year} {2012})}\BibitemShut {NoStop}%
\bibitem [{\citenamefont {Schliwa}\ and\ \citenamefont
  {Woehlke}(2003)}]{Schliwa_Nature_2003}%
  \BibitemOpen
  \bibfield  {author} {\bibinfo {author} {\bibfnamefont {M.}~\bibnamefont
  {Schliwa}}\ and\ \bibinfo {author} {\bibfnamefont {G.}~\bibnamefont
  {Woehlke}},\ }\href@noop {} {\bibfield  {journal} {\bibinfo  {journal}
  {Nature}\ }\textbf {\bibinfo {volume} {422}},\ \bibinfo {pages} {759}
  (\bibinfo {year} {2003})}\BibitemShut {NoStop}%
\bibitem [{\citenamefont {Nagai}\ \emph {et~al.}(2005)\citenamefont {Nagai},
  \citenamefont {Sumino}, \citenamefont {Kitahata},\ and\ \citenamefont
  {Yoshikawa}}]{Nagai_PRE_2005}%
  \BibitemOpen
  \bibfield  {author} {\bibinfo {author} {\bibfnamefont {K.}~\bibnamefont
  {Nagai}}, \bibinfo {author} {\bibfnamefont {Y.}~\bibnamefont {Sumino}},
  \bibinfo {author} {\bibfnamefont {H.}~\bibnamefont {Kitahata}}, \ and\
  \bibinfo {author} {\bibfnamefont {K.}~\bibnamefont {Yoshikawa}},\ }\href@noop
  {} {\bibfield  {journal} {\bibinfo  {journal} {Phys. Rev. E}\ }\textbf
  {\bibinfo {volume} {71}},\ \bibinfo {pages} {065301(R)} (\bibinfo {year}
  {2005})}\BibitemShut {NoStop}%
\bibitem [{\citenamefont {Yabunaka}\ \emph {et~al.}(2012)\citenamefont
  {Yabunaka}, \citenamefont {Yoshinaga},\ and\ \citenamefont
  {Ohta}}]{Yabunaka_JCP_2012}%
  \BibitemOpen
  \bibfield  {author} {\bibinfo {author} {\bibfnamefont {S.}~\bibnamefont
  {Yabunaka}}, \bibinfo {author} {\bibfnamefont {N.}~\bibnamefont {Yoshinaga}},
  \ and\ \bibinfo {author} {\bibfnamefont {T.}~\bibnamefont {Ohta}},\
  }\href@noop {} {\bibfield  {journal} {\bibinfo  {journal} {J. Chem. Phys.}\
  }\textbf {\bibinfo {volume} {136}},\ \bibinfo {pages} {074904} (\bibinfo
  {year} {2012})}\BibitemShut {NoStop}%
\bibitem [{\citenamefont {Golestanian}\ \emph {et~al.}(2007)\citenamefont
  {Golestanian}, \citenamefont {Liverpool},\ and\ \citenamefont
  {Ajdari}}]{Golestanian_NJP_2007}%
  \BibitemOpen
  \bibfield  {author} {\bibinfo {author} {\bibfnamefont {R.}~\bibnamefont
  {Golestanian}}, \bibinfo {author} {\bibfnamefont {T.~B.}\ \bibnamefont
  {Liverpool}}, \ and\ \bibinfo {author} {\bibfnamefont {A.}~\bibnamefont
  {Ajdari}},\ }\href@noop {} {\bibfield  {journal} {\bibinfo  {journal} {New J.
  Phys.}\ }\textbf {\bibinfo {volume} {9}},\ \bibinfo {pages} {126} (\bibinfo
  {year} {2007})}\BibitemShut {NoStop}%
\bibitem [{\citenamefont {Ebbens}\ and\ \citenamefont
  {Howse}(2010)}]{Ebbens_SM_2010}%
  \BibitemOpen
  \bibfield  {author} {\bibinfo {author} {\bibfnamefont {S.~J.}\ \bibnamefont
  {Ebbens}}\ and\ \bibinfo {author} {\bibfnamefont {J.~R.}\ \bibnamefont
  {Howse}},\ }\href@noop {} {\bibfield  {journal} {\bibinfo  {journal} {Soft
  Matter}\ }\textbf {\bibinfo {volume} {6}},\ \bibinfo {pages} {726} (\bibinfo
  {year} {2010})}\BibitemShut {NoStop}%
\bibitem [{\citenamefont {Golestanian}\ \emph {et~al.}(2005)\citenamefont
  {Golestanian}, \citenamefont {Liverpool},\ and\ \citenamefont
  {Ajdari}}]{Golestanian_PRL_2005}%
  \BibitemOpen
  \bibfield  {author} {\bibinfo {author} {\bibfnamefont {R.}~\bibnamefont
  {Golestanian}}, \bibinfo {author} {\bibfnamefont {T.~B.}\ \bibnamefont
  {Liverpool}}, \ and\ \bibinfo {author} {\bibfnamefont {A.}~\bibnamefont
  {Ajdari}},\ }\href@noop {} {\bibfield  {journal} {\bibinfo  {journal} {Phys.
  Rev. Lett.}\ }\textbf {\bibinfo {volume} {94}},\ \bibinfo {pages} {220801}
  (\bibinfo {year} {2005})}\BibitemShut {NoStop}%
\bibitem [{\citenamefont {Reddy}\ and\ \citenamefont
  {Clasen}(2014)}]{Reddy_KARJ_2014}%
  \BibitemOpen
  \bibfield  {author} {\bibinfo {author} {\bibfnamefont {N.~K.}\ \bibnamefont
  {Reddy}}\ and\ \bibinfo {author} {\bibfnamefont {C.}~\bibnamefont {Clasen}},\
  }\href@noop {} {\bibfield  {journal} {\bibinfo  {journal} {Korea-Aust. Rheol.
  J.}\ }\textbf {\bibinfo {volume} {26}},\ \bibinfo {pages} {73} (\bibinfo
  {year} {2014})}\BibitemShut {NoStop}%
\bibitem [{\citenamefont {Anderson}(1989)}]{Anderson_ARFM_1989}%
  \BibitemOpen
  \bibfield  {author} {\bibinfo {author} {\bibfnamefont {J.~L.}\ \bibnamefont
  {Anderson}},\ }\href@noop {} {\bibfield  {journal} {\bibinfo  {journal}
  {Annu. Rev. Fluid Mech.}\ }\textbf {\bibinfo {volume} {21}},\ \bibinfo
  {pages} {61} (\bibinfo {year} {1989})}\BibitemShut {NoStop}%
\bibitem [{\citenamefont {Gangwal}\ \emph {et~al.}(2008)\citenamefont
  {Gangwal}, \citenamefont {Cayre}, \citenamefont {Bazant},\ and\ \citenamefont
  {Velev}}]{Gangwal_PRL_2008}%
  \BibitemOpen
  \bibfield  {author} {\bibinfo {author} {\bibfnamefont {S.}~\bibnamefont
  {Gangwal}}, \bibinfo {author} {\bibfnamefont {O.~J.}\ \bibnamefont {Cayre}},
  \bibinfo {author} {\bibfnamefont {M.~Z.}\ \bibnamefont {Bazant}}, \ and\
  \bibinfo {author} {\bibfnamefont {O.~D.}\ \bibnamefont {Velev}},\ }\href@noop
  {} {\bibfield  {journal} {\bibinfo  {journal} {Phys. Rev. Lett.}\ }\textbf
  {\bibinfo {volume} {100}},\ \bibinfo {pages} {058302} (\bibinfo {year}
  {2008})}\BibitemShut {NoStop}%
\bibitem [{\citenamefont {Volpe}\ \emph {et~al.}(2011)\citenamefont {Volpe},
  \citenamefont {Buttinoni}, \citenamefont {Vogt}, \citenamefont
  {K\"{u}mmerer},\ and\ \citenamefont {Bechinger}}]{Volpe_SM_2011}%
  \BibitemOpen
  \bibfield  {author} {\bibinfo {author} {\bibfnamefont {G.}~\bibnamefont
  {Volpe}}, \bibinfo {author} {\bibfnamefont {I.}~\bibnamefont {Buttinoni}},
  \bibinfo {author} {\bibfnamefont {D.}~\bibnamefont {Vogt}}, \bibinfo {author}
  {\bibfnamefont {H.-J.}\ \bibnamefont {K\"{u}mmerer}}, \ and\ \bibinfo
  {author} {\bibfnamefont {C.}~\bibnamefont {Bechinger}},\ }\href@noop {}
  {\bibfield  {journal} {\bibinfo  {journal} {Soft Matter}\ }\textbf {\bibinfo
  {volume} {7}},\ \bibinfo {pages} {8810} (\bibinfo {year} {2011})}\BibitemShut
  {NoStop}%
\bibitem [{\citenamefont {Buttinoni}\ \emph {et~al.}(2012)\citenamefont
  {Buttinoni}, \citenamefont {Volpe}, \citenamefont {K\"{u}mmel}, \citenamefont
  {Volpe},\ and\ \citenamefont {Bechinger}}]{Buttinoni_JPCM_2012}%
  \BibitemOpen
  \bibfield  {author} {\bibinfo {author} {\bibfnamefont {I.}~\bibnamefont
  {Buttinoni}}, \bibinfo {author} {\bibfnamefont {G.}~\bibnamefont {Volpe}},
  \bibinfo {author} {\bibfnamefont {F.}~\bibnamefont {K\"{u}mmel}}, \bibinfo
  {author} {\bibfnamefont {G.}~\bibnamefont {Volpe}}, \ and\ \bibinfo {author}
  {\bibfnamefont {C.}~\bibnamefont {Bechinger}},\ }\href@noop {} {\bibfield
  {journal} {\bibinfo  {journal} {J. Phys.: Condens. Matter}\ }\textbf
  {\bibinfo {volume} {24}},\ \bibinfo {pages} {284129} (\bibinfo {year}
  {2012})}\BibitemShut {NoStop}%
\bibitem [{\citenamefont {Cahn}(1977)}]{Cahn_JCP_1977}%
  \BibitemOpen
  \bibfield  {author} {\bibinfo {author} {\bibfnamefont {J.~W.}\ \bibnamefont
  {Cahn}},\ }\href@noop {} {\bibfield  {journal} {\bibinfo  {journal} {J. Chem.
  Phys.}\ }\textbf {\bibinfo {volume} {66}},\ \bibinfo {pages} {3667} (\bibinfo
  {year} {1977})}\BibitemShut {NoStop}%
\bibitem [{\citenamefont {de~Gennes}(1985)}]{deGennes_RPM_1985}%
  \BibitemOpen
  \bibfield  {author} {\bibinfo {author} {\bibfnamefont {P.~G.}\ \bibnamefont
  {de~Gennes}},\ }\href@noop {} {\bibfield  {journal} {\bibinfo  {journal}
  {Rev. Mod. Phys.}\ }\textbf {\bibinfo {volume} {57}},\ \bibinfo {pages} {827}
  (\bibinfo {year} {1985})}\BibitemShut {NoStop}%
\bibitem [{\citenamefont {Binder}\ and\ \citenamefont
  {Stauffer}(1974)}]{Binder_PRL_1974}%
  \BibitemOpen
  \bibfield  {author} {\bibinfo {author} {\bibfnamefont {K.}~\bibnamefont
  {Binder}}\ and\ \bibinfo {author} {\bibfnamefont {D.}~\bibnamefont
  {Stauffer}},\ }\href@noop {} {\bibfield  {journal} {\bibinfo  {journal}
  {Phys. Rev. Lett.}\ }\textbf {\bibinfo {volume} {33}},\ \bibinfo {pages}
  {1006} (\bibinfo {year} {1974})}\BibitemShut {NoStop}%
\bibitem [{\citenamefont {Siggia}(1979)}]{Siggia_PRA_1979}%
  \BibitemOpen
  \bibfield  {author} {\bibinfo {author} {\bibfnamefont {E.~D.}\ \bibnamefont
  {Siggia}},\ }\href@noop {} {\bibfield  {journal} {\bibinfo  {journal} {Phys.
  Rev. A}\ }\textbf {\bibinfo {volume} {20}},\ \bibinfo {pages} {595} (\bibinfo
  {year} {1979})}\BibitemShut {NoStop}%
\bibitem [{\citenamefont {Onuki}(2002)}]{Onuki_book_2002}%
  \BibitemOpen
  \bibfield  {author} {\bibinfo {author} {\bibfnamefont {A.}~\bibnamefont
  {Onuki}},\ }\href@noop {} {\emph {\bibinfo {title} {Phase Transition
  Dynamics}}}\ (\bibinfo  {publisher} {Cambridge Unversity Press, Cambridge},\
  \bibinfo {year} {2002})\BibitemShut {NoStop}%
\bibitem [{\citenamefont {Tanaka}\ \emph {et~al.}(1994)\citenamefont {Tanaka},
  \citenamefont {Lovinger},\ and\ \citenamefont {Davis}}]{Tanaka_PRL_1984}%
  \BibitemOpen
  \bibfield  {author} {\bibinfo {author} {\bibfnamefont {H.}~\bibnamefont
  {Tanaka}}, \bibinfo {author} {\bibfnamefont {A.~J.}\ \bibnamefont
  {Lovinger}}, \ and\ \bibinfo {author} {\bibfnamefont {D.~D.}\ \bibnamefont
  {Davis}},\ }\href@noop {} {\bibfield  {journal} {\bibinfo  {journal} {Phys
  Rev. Lett.}\ }\textbf {\bibinfo {volume} {72}},\ \bibinfo {pages} {2581}
  (\bibinfo {year} {1994})}\BibitemShut {NoStop}%
\bibitem [{\citenamefont {Ginzburg}\ \emph {et~al.}(1999)\citenamefont
  {Ginzburg}, \citenamefont {Peng}, \citenamefont {Qiu}, \citenamefont
  {Jasnow},\ and\ \citenamefont {Balazs}}]{Ginzburg_PRE_1999}%
  \BibitemOpen
  \bibfield  {author} {\bibinfo {author} {\bibfnamefont {V.~V.}\ \bibnamefont
  {Ginzburg}}, \bibinfo {author} {\bibfnamefont {G.}~\bibnamefont {Peng}},
  \bibinfo {author} {\bibfnamefont {F.}~\bibnamefont {Qiu}}, \bibinfo {author}
  {\bibfnamefont {D.}~\bibnamefont {Jasnow}}, \ and\ \bibinfo {author}
  {\bibfnamefont {A.~C.}\ \bibnamefont {Balazs}},\ }\href@noop {} {\bibfield
  {journal} {\bibinfo  {journal} {Phys Rev. E}\ }\textbf {\bibinfo {volume}
  {60}},\ \bibinfo {pages} {4352} (\bibinfo {year} {1999})}\BibitemShut
  {NoStop}%
\bibitem [{\citenamefont {Araki}\ and\ \citenamefont
  {Tanaka}(2006)}]{Araki_PRE_2006}%
  \BibitemOpen
  \bibfield  {author} {\bibinfo {author} {\bibfnamefont {T.}~\bibnamefont
  {Araki}}\ and\ \bibinfo {author} {\bibfnamefont {H.}~\bibnamefont {Tanaka}},\
  }\href@noop {} {\bibfield  {journal} {\bibinfo  {journal} {Phys. Rev. E}\
  }\textbf {\bibinfo {volume} {73}},\ \bibinfo {pages} {061506} (\bibinfo
  {year} {2006})}\BibitemShut {NoStop}%
\bibitem [{\citenamefont {Cates}\ and\ \citenamefont
  {Clegg}(2008)}]{Cates_SM_2008}%
  \BibitemOpen
  \bibfield  {author} {\bibinfo {author} {\bibfnamefont {M.~E.}\ \bibnamefont
  {Cates}}\ and\ \bibinfo {author} {\bibfnamefont {P.~S.}\ \bibnamefont
  {Clegg}},\ }\href@noop {} {\bibfield  {journal} {\bibinfo  {journal} {Soft
  Matter}\ }\textbf {\bibinfo {volume} {4}},\ \bibinfo {pages} {2132} (\bibinfo
  {year} {2008})}\BibitemShut {NoStop}%
\bibitem [{\citenamefont {Binks}\ and\ \citenamefont
  {Fletcher}(2001)}]{Binks_Lang_2001}%
  \BibitemOpen
  \bibfield  {author} {\bibinfo {author} {\bibfnamefont {B.~P.}\ \bibnamefont
  {Binks}}\ and\ \bibinfo {author} {\bibfnamefont {P.~D.~I.}\ \bibnamefont
  {Fletcher}},\ }\href@noop {} {\bibfield  {journal} {\bibinfo  {journal}
  {Langmuir}\ }\textbf {\bibinfo {volume} {17}},\ \bibinfo {pages} {4708}
  (\bibinfo {year} {2001})}\BibitemShut {NoStop}%
\bibitem [{\citenamefont {Glaser}\ \emph {et~al.}(2006)\citenamefont {Glaser},
  \citenamefont {Adams}, \citenamefont {B\`{o}ker},\ and\ \citenamefont
  {Krausch}}]{Glaser_Lang_2006}%
  \BibitemOpen
  \bibfield  {author} {\bibinfo {author} {\bibfnamefont {N.}~\bibnamefont
  {Glaser}}, \bibinfo {author} {\bibfnamefont {D.~J.}\ \bibnamefont {Adams}},
  \bibinfo {author} {\bibfnamefont {A.}~\bibnamefont {B\`{o}ker}}, \ and\
  \bibinfo {author} {\bibfnamefont {G.}~\bibnamefont {Krausch}},\ }\href@noop
  {} {\bibfield  {journal} {\bibinfo  {journal} {Langmuir}\ }\textbf {\bibinfo
  {volume} {22}},\ \bibinfo {pages} {5227} (\bibinfo {year}
  {2006})}\BibitemShut {NoStop}%
\bibitem [{\citenamefont {Huang}\ \emph {et~al.}(2012)\citenamefont {Huang},
  \citenamefont {Li},\ and\ \citenamefont {Guo}}]{Huang_SM_2012}%
  \BibitemOpen
  \bibfield  {author} {\bibinfo {author} {\bibfnamefont {M.}~\bibnamefont
  {Huang}}, \bibinfo {author} {\bibfnamefont {Z.}~\bibnamefont {Li}}, \ and\
  \bibinfo {author} {\bibfnamefont {H.}~\bibnamefont {Guo}},\ }\href@noop {}
  {\bibfield  {journal} {\bibinfo  {journal} {Soft Matter}\ }\textbf {\bibinfo
  {volume} {8}},\ \bibinfo {pages} {6834} (\bibinfo {year} {2012})}\BibitemShut
  {NoStop}%
\bibitem [{\citenamefont {Onuki}(1982{\natexlab{a}})}]{Onuki_PTP_1982}%
  \BibitemOpen
  \bibfield  {author} {\bibinfo {author} {\bibfnamefont {A.}~\bibnamefont
  {Onuki}},\ }\href@noop {} {\bibfield  {journal} {\bibinfo  {journal} {Prog.
  Theor. Phys.}\ }\textbf {\bibinfo {volume} {67}},\ \bibinfo {pages} {1740}
  (\bibinfo {year} {1982}{\natexlab{a}})}\BibitemShut {NoStop}%
\bibitem [{\citenamefont {Onuki}(1982{\natexlab{b}})}]{Onuki_PRL_1982}%
  \BibitemOpen
  \bibfield  {author} {\bibinfo {author} {\bibfnamefont {A.}~\bibnamefont
  {Onuki}},\ }\href@noop {} {\bibfield  {journal} {\bibinfo  {journal} {Phys.
  Rev. Lett.}\ }\textbf {\bibinfo {volume} {48}},\ \bibinfo {pages} {753}
  (\bibinfo {year} {1982}{\natexlab{b}})}\BibitemShut {NoStop}%
\bibitem [{\citenamefont {Joshua}\ \emph {et~al.}(1985)\citenamefont {Joshua},
  \citenamefont {Goldburg},\ and\ \citenamefont {Onuki}}]{Joshua_PRL_1985}%
  \BibitemOpen
  \bibfield  {author} {\bibinfo {author} {\bibfnamefont {M.}~\bibnamefont
  {Joshua}}, \bibinfo {author} {\bibfnamefont {W.~I.}\ \bibnamefont
  {Goldburg}}, \ and\ \bibinfo {author} {\bibfnamefont {A.}~\bibnamefont
  {Onuki}},\ }\href@noop {} {\bibfield  {journal} {\bibinfo  {journal} {Phys
  Rev. Lett.}\ }\textbf {\bibinfo {volume} {54}},\ \bibinfo {pages} {1175}
  (\bibinfo {year} {1985})}\BibitemShut {NoStop}%
\bibitem [{\citenamefont {Tanaka}\ and\ \citenamefont
  {Sigehuzi}(1995)}]{Tanaka_PRL_1995}%
  \BibitemOpen
  \bibfield  {author} {\bibinfo {author} {\bibfnamefont {H.}~\bibnamefont
  {Tanaka}}\ and\ \bibinfo {author} {\bibfnamefont {T.}~\bibnamefont
  {Sigehuzi}},\ }\href@noop {} {\bibfield  {journal} {\bibinfo  {journal}
  {Phys. Rev. Lett.}\ }\textbf {\bibinfo {volume} {75}},\ \bibinfo {pages}
  {874} (\bibinfo {year} {1995})}\BibitemShut {NoStop}%
\bibitem [{\citenamefont {Tanaka}\ and\ \citenamefont
  {Araki}(2000{\natexlab{a}})}]{Tanaka_PRL_2000}%
  \BibitemOpen
  \bibfield  {author} {\bibinfo {author} {\bibfnamefont {H.}~\bibnamefont
  {Tanaka}}\ and\ \bibinfo {author} {\bibfnamefont {T.}~\bibnamefont {Araki}},\
  }\href@noop {} {\bibfield  {journal} {\bibinfo  {journal} {Phys. Rev. Lett.}\
  }\textbf {\bibinfo {volume} {85}},\ \bibinfo {pages} {1338} (\bibinfo {year}
  {2000}{\natexlab{a}})}\BibitemShut {NoStop}%
\bibitem [{\citenamefont {Tanaka}\ and\ \citenamefont
  {Araki}(2006)}]{Tanaka_CES_2006}%
  \BibitemOpen
  \bibfield  {author} {\bibinfo {author} {\bibfnamefont {H.}~\bibnamefont
  {Tanaka}}\ and\ \bibinfo {author} {\bibfnamefont {T.}~\bibnamefont {Araki}},\
  }\href@noop {} {\bibfield  {journal} {\bibinfo  {journal} {Chem. Eng. Sci.}\
  }\textbf {\bibinfo {volume} {61}},\ \bibinfo {pages} {2108} (\bibinfo {year}
  {2006})}\BibitemShut {NoStop}%
\bibitem [{\citenamefont {Araki}\ and\ \citenamefont
  {Tanaka}(2008)}]{Araki_JPCM_2008}%
  \BibitemOpen
  \bibfield  {author} {\bibinfo {author} {\bibfnamefont {T.}~\bibnamefont
  {Araki}}\ and\ \bibinfo {author} {\bibfnamefont {H.}~\bibnamefont {Tanaka}},\
  }\href@noop {} {\bibfield  {journal} {\bibinfo  {journal} {J. Phys.: Condens.
  Matter}\ }\textbf {\bibinfo {volume} {20}},\ \bibinfo {pages} {072101}
  (\bibinfo {year} {2008})}\BibitemShut {NoStop}%
\bibitem [{\citenamefont {Furukawa}\ \emph {et~al.}(2013)\citenamefont
  {Furukawa}, \citenamefont {Gambassi}, \citenamefont {Dietrich},\ and\
  \citenamefont {Tanaka}}]{Furukawa_PRL_2013}%
  \BibitemOpen
  \bibfield  {author} {\bibinfo {author} {\bibfnamefont {A.}~\bibnamefont
  {Furukawa}}, \bibinfo {author} {\bibfnamefont {A.}~\bibnamefont {Gambassi}},
  \bibinfo {author} {\bibfnamefont {S.}~\bibnamefont {Dietrich}}, \ and\
  \bibinfo {author} {\bibfnamefont {H.}~\bibnamefont {Tanaka}},\ }\href@noop {}
  {\bibfield  {journal} {\bibinfo  {journal} {Phys. Rev. Lett.}\ }\textbf
  {\bibinfo {volume} {111}},\ \bibinfo {pages} {055701} (\bibinfo {year}
  {2013})}\BibitemShut {NoStop}%
\bibitem [{\citenamefont {Harlow}\ and\ \citenamefont
  {Welch}(1965)}]{Harlow_PF_1965}%
  \BibitemOpen
  \bibfield  {author} {\bibinfo {author} {\bibfnamefont {F.~H.}\ \bibnamefont
  {Harlow}}\ and\ \bibinfo {author} {\bibfnamefont {J.~D.}\ \bibnamefont
  {Welch}},\ }\href@noop {} {\bibfield  {journal} {\bibinfo  {journal} {Phys.
  Fluids.}\ }\textbf {\bibinfo {volume} {8}},\ \bibinfo {pages} {2182}
  (\bibinfo {year} {1965})}\BibitemShut {NoStop}%
\bibitem [{\citenamefont {Tanaka}\ and\ \citenamefont
  {Araki}(1998)}]{Tanaka_PRL_1998}%
  \BibitemOpen
  \bibfield  {author} {\bibinfo {author} {\bibfnamefont {H.}~\bibnamefont
  {Tanaka}}\ and\ \bibinfo {author} {\bibfnamefont {T.}~\bibnamefont {Araki}},\
  }\href@noop {} {\bibfield  {journal} {\bibinfo  {journal} {Phys. Rev. Lett.}\
  }\textbf {\bibinfo {volume} {81}},\ \bibinfo {pages} {389} (\bibinfo {year}
  {1998})}\BibitemShut {NoStop}%
\bibitem [{\citenamefont {Tanaka}\ and\ \citenamefont
  {Araki}(2000{\natexlab{b}})}]{Tanaka_EPL_2000}%
  \BibitemOpen
  \bibfield  {author} {\bibinfo {author} {\bibfnamefont {H.}~\bibnamefont
  {Tanaka}}\ and\ \bibinfo {author} {\bibfnamefont {T.}~\bibnamefont {Araki}},\
  }\href@noop {} {\bibfield  {journal} {\bibinfo  {journal} {Europhys. Lett.}\
  }\textbf {\bibinfo {volume} {51}},\ \bibinfo {pages} {154} (\bibinfo {year}
  {2000}{\natexlab{b}})}\BibitemShut {NoStop}%
\bibitem [{\citenamefont {Karim}\ \emph {et~al.}(1999)\citenamefont {Karim},
  \citenamefont {Douglas}, \citenamefont {Nisato}, \citenamefont {Liu},\ and\
  \citenamefont {Amis}}]{Karim_Macro_1999}%
  \BibitemOpen
  \bibfield  {author} {\bibinfo {author} {\bibfnamefont {A.}~\bibnamefont
  {Karim}}, \bibinfo {author} {\bibfnamefont {J.~F.}\ \bibnamefont {Douglas}},
  \bibinfo {author} {\bibfnamefont {G.}~\bibnamefont {Nisato}}, \bibinfo
  {author} {\bibfnamefont {D.-W.}\ \bibnamefont {Liu}}, \ and\ \bibinfo
  {author} {\bibfnamefont {E.~J.}\ \bibnamefont {Amis}},\ }\href@noop {}
  {\bibfield  {journal} {\bibinfo  {journal} {Macromolecules}\ }\textbf
  {\bibinfo {volume} {32}},\ \bibinfo {pages} {5917} (\bibinfo {year}
  {1999})}\BibitemShut {NoStop}%
\bibitem [{\citenamefont {Winter}\ \emph {et~al.}(2009)\citenamefont {Winter},
  \citenamefont {Virnau},\ and\ \citenamefont {BInder}}]{Winter_PRL_2009}%
  \BibitemOpen
  \bibfield  {author} {\bibinfo {author} {\bibfnamefont {D.}~\bibnamefont
  {Winter}}, \bibinfo {author} {\bibfnamefont {P.}~\bibnamefont {Virnau}}, \
  and\ \bibinfo {author} {\bibfnamefont {K.}~\bibnamefont {BInder}},\
  }\href@noop {} {\bibfield  {journal} {\bibinfo  {journal} {Phys. Rev. Lett.}\
  }\textbf {\bibinfo {volume} {103}},\ \bibinfo {pages} {225703} (\bibinfo
  {year} {2009})}\BibitemShut {NoStop}%
\bibitem [{\citenamefont {Shibayama}\ \emph {et~al.}(2004)\citenamefont
  {Shibayama}, \citenamefont {Isono}, \citenamefont {Okabe}, \citenamefont
  {Karino},\ and\ \citenamefont {Nagao}}]{Shibayama_Macro_2004}%
  \BibitemOpen
  \bibfield  {author} {\bibinfo {author} {\bibfnamefont {M.}~\bibnamefont
  {Shibayama}}, \bibinfo {author} {\bibfnamefont {K.}~\bibnamefont {Isono}},
  \bibinfo {author} {\bibfnamefont {S.}~\bibnamefont {Okabe}}, \bibinfo
  {author} {\bibfnamefont {T.}~\bibnamefont {Karino}}, \ and\ \bibinfo {author}
  {\bibfnamefont {M.}~\bibnamefont {Nagao}},\ }\href@noop {} {\bibfield
  {journal} {\bibinfo  {journal} {Macromolecules}\ }\textbf {\bibinfo {volume}
  {37}},\ \bibinfo {pages} {2909} (\bibinfo {year} {2004})}\BibitemShut
  {NoStop}%
\end{thebibliography}%
\bibliographystyle{apsrev4-1} 

\appendix
\section{Time development of total free energy}
\label{sec:app1}

The total free energy including the kinetic energy is considered. 
Using Eq.~(\ref{eq:F}), it is given by
\bea
\mathcal{G}\{\phi,\bi{v},\bi{R},\bi{n}\}
=\mathcal{F}+\frac{1}{2}\int d\bi{r}\rho \bi{v}^2. 
\ena
Its time development is 
\bea
\frac{d}{dt}\mathcal {G}&=&\int d\bi{r}\left[
\frac{\delta \mathcal{F}}{\delta\phi}\dot{\phi}+\rho \dot{\bi{v}}\cdot\bi{v}
\right]\nonumber\\
&&+\left\{\frac{\partial \mathcal{F}}{\partial \bi{R}}\cdot \dot{\bi{R}}+\frac{\partial \mathcal{F}}{\partial \bi{n}}\cdot \dot{\bi{n}}\right\}. 
\ena
After some calculations, 
we obtain 
\bea
\frac{d}{dt}\mathcal{G}
&=&\int d\bi{r}\left\{-L\left(\n\frac{\partial \mathcal{F}}{\delta \phi}\right)^2
-\n \bi{v}:\bi{\Sigma}+\bi{f}\cdot\bi{v}
\right.\nonumber\\
&&\left.
+ \bi{g}\cdot(\n\times \bi{v})
+\left(p+\frac{1}{2}\rho\bi{v}^2\right)\n \cdot\bi{v}\right\}\nonumber\\
&&+\left\{\frac{\partial \mathcal{F}}{\partial \bi{R}}\cdot \dot{\bi{R}}+\frac{\partial \mathcal{F}}{\partial \bi{n}}\cdot \dot{\bi{n}}\right\}, 
\ena
where we neglect the contribution of the system boundary.  
For the force $\bi{f}$ and torque $\bi{g}$ fields, 
we obtained the following equations in our scheme. 
\bea
\int d\bi{r}\bi{f}\cdot \bi{v}&=&
-\frac{1}{V}\int d\bi{r}
\psi(\bi{r}-\bi{R})\frac{\partial\mathcal{F}}{\partial \bi{R}}\cdot 
\bi{v}(\bi{r})
\nonumber\\
&=&-\frac{1}{V} \frac{\partial \mathcal{F}}{\partial \bi{R}}\cdot\int d\bi{r}\psi(\bi{r}-\bi{R})\bi{v}(\bi{r})\nonumber\\
&=&-\frac{\partial \mathcal{F}}{\partial \bi{R}}\cdot \dot{\bi{R}},
\ena
\bea
&&\int d\bi{r}\bi{g}\cdot (\n\times \bi{v})\nonumber\\
&=&- \int d\bi{r}\frac{\psi(\bi{r}-\bi{R})}{V}\left\{\bi{n}\times \frac{\partial \mathcal{F}}{\partial \bi{n}}\right\}\cdot \{\n\times\bi{v}(\bi{r})\}\nonumber\\
&=&-\frac{\partial \mathcal{F}}{\partial\bi{n}} \cdot \left[
\frac{1}{V}\int d\bi{r}\psi(\bi{r}-\bi{R})\{\n\times \bi{v}(\bi{r})\}\times\bi{n}\right]
\nonumber\\
&=&-\int d\bi{r}\frac{\partial\mathcal{F}}{\partial \bi{n}}\cdot\dot{\bi{n}}
\ena
Then, we finally obtain 
\bea
\frac{d}{dt}\mathcal{G}=-\int d\bi{r}\left\{
L\left(\n\frac{\delta \mathcal{F}}{\delta \phi}\right)^2+\frac{1}{2}\eta
\{\n\bi{v}+(\n\bi{v})^T\}^2\right\}, \nonumber\\
\ena
where we assumed $\n\cdot \bi{v}=0$. 
Thus, we confirmed that $\mathcal{G}$ decreases with time only 
via diffusion and viscous dissipation.

\end{document}